\documentclass[aps,prd,nofootinbib,superscriptaddress,10pt]{revtex4}
\usepackage{txfonts}
\usepackage{graphicx}
\usepackage{dcolumn}
\usepackage{bm}
\usepackage{amssymb}
\usepackage{latexsym}
\usepackage[colorlinks, linkcolor=blue, citecolor=blue, urlcolor=blue]{hyperref}

\newcommand{\be}{\begin{equation}}
\newcommand{\ee}{\end{equation}}
\newcommand{\bq}{\begin{eqnarray}}
\newcommand{\eq}{\end{eqnarray}}

\begin{document}

\title{Probing interaction and spatial curvature in the holographic dark energy model}

\author{Miao Li}
\email{mli@itp.ac.cn} \affiliation{Kavli Institute for Theoretical
Physics China, Chinese Academy of Sciences, Beijing 100190, China}
\affiliation{Key Laboratory of Frontiers in Theoretical Physics,
Institute of Theoretical Physics, Chinese Academy of Sciences,
Beijing 100190, China}

\author{Xiao-Dong Li}
\email{renzhe@mail.ustc.edu.cn} \affiliation{Interdisciplinary
Center for Theoretical Study, University of Science and Technology
of China, Hefei 230026, China} \affiliation{Key Laboratory of
Frontiers in Theoretical Physics, Institute of Theoretical Physics,
Chinese Academy of Sciences, Beijing 100190, China}

\author{Shuang Wang}
\email{swang@mail.ustc.edu.cn} \affiliation{Department of Modern
Physics, University of Science and Technology of China, Hefei
230026, China} \affiliation{Key Laboratory of Frontiers in
Theoretical Physics, Institute of Theoretical Physics, Chinese
Academy of Sciences, Beijing 100190, China}

\author{Yi Wang}
\email{wangyi@hep.physics.mcgill.ca} \affiliation{Physics
Department, McGill University, Montreal, H3A2T8, Canada}

\author{Xin Zhang}
\email{zhangxin@mail.neu.edu.cn} \affiliation{Department of Physics,
College of Sciences, Northeastern University, Shenyang 110004,
China} \affiliation{Kavli Institute for Theoretical Physics China,
Chinese Academy of Sciences, Beijing 100190, China}

\begin{abstract}
In this paper we place observational constraints on the interaction
and spatial curvature in the holographic dark energy model. We
consider three kinds of phenomenological interactions between
holographic dark energy and matter, i.e., the interaction term $Q$
is proportional to the energy densities of dark energy
($\rho_{\Lambda}$), matter ($\rho_{m}$), and matter plus dark energy
($\rho_m+\rho_{\Lambda}$). For probing the interaction and spatial
curvature in the holographic dark energy model, we use the latest
observational data including the type Ia supernovae (SNIa)
Constitution data, the shift parameter of the cosmic microwave
background (CMB) given by the five-year Wilkinson Microwave
Anisotropy Probe (WMAP5) observations, and the baryon acoustic
oscillation (BAO) measurement from the Sloan Digital Sky Survey
(SDSS). Our results show that the interaction and spatial curvature
in the holographic dark energy model are both rather small. Besides,
it is interesting to find that there exists significant degeneracy
between the phenomenological interaction and the spatial curvature
in the holographic dark energy model.
\end{abstract}

\maketitle

\section{Introduction}
Observations of type Ia supernovae (SNIa) \cite{Riess}, cosmic
microwave background (CMB) \cite{spergel} and large scale structure
(LSS) \cite{Tegmark} all indicate the existence of mysterious dark
energy driving the current accelerated expansion of universe, and
lots of efforts have been made to understand it. The most important
theoretical candidate of dark energy is the Einstein's cosmological
constant $\lambda$, which can fit the observations well so far, but
is plagued with the famous fine-tuning and cosmic coincidence
problems \cite{Weinberg}. Many other dynamical dark energy models
have also been proposed in the literature, such as quintessence
\cite{quint}, phantom \cite{phantom}, $k$-essence \cite{k}, tachyon
\cite{tachyonic}, hessence \cite{hessence}, Chaplygin gas
\cite{Chaplygin}, and Yang-Mills condensate \cite{YMC}, etc.

Actually, the dark energy problem may be in essence an issue of
quantum gravity \cite{Witten:2000zk}. However, by far, we have no a
complete theory of quantum gravity, so it seems that we have to
consider the effects of gravity in some effective quantum field
theory in which some fundamental principles of quantum gravity
should be taken into account. It is commonly believed that the
holographic principle \cite{Holography} is just a fundamental
principle of quantum gravity. Based on the effective quantum field
theory, Cohen et al. \cite{cohen99} pointed out that the quantum
zero-point energy of a system with size $L$ should not exceed the
mass of a black hole with the same size, i.e. $L^3\Lambda^4\leq
LM_{Pl}^2$, where $\Lambda$ is the ultraviolet (UV) cutoff of the
effective quantum field theory, which is closely related to the
quantum zero-point energy density, and $M_{Pl}\equiv 1/\sqrt{8\pi
G}$ is the reduced Planck mass. This observation relates the UV
cutoff of a system to its infrared (IR) cutoff. When we take the
whole universe into account, the vacuum energy related to this
holographic principle can be viewed as dark energy (its energy
density is denoted as $\rho_{\Lambda}$ hereafter). The largest IR
cutoff $L$ is chosen by saturating the inequality, so that we get
the holographic dark energy density
\begin{equation}
\label{eq:rhode0}\rho_{\Lambda}=3c^2M_{Pl}^2L^{-2}
\end{equation}
where $c$ is a numerical constant characterizing all of the
uncertainties of the theory, and its value can only be determined by
observations. If we take $L$ as the size of the current universe,
say, the Hubble radius $H^{-1}$, then the dark energy density will
be close to the observational result. However, Hsu \cite{hsu04}
pointed out this yields a wrong equation of state for dark energy.
Subsequently, Li \cite{li04} suggested to choose the future event
horizon of the universe as the IR cutoff of this theory. This choice
not only gives a reasonable value for dark energy, but also leads to
an accelerated universe. Moreover, the cosmic coincidence problem
can also be explained successfully in this model. Most recently, a
calculation of the Casimir energy of the photon field in a de Sitter
space is performed \cite{lmp}, and it is a surprising result that
the Casimir energy is indeed proportional to the size of the horizon
(the usual Casimir energy in a cavity is inversely proportional to
the size of the cavity), in agreement with the holographic dark
energy model.

The holographic dark energy model has been studied widely
\cite{holode1,holode2,holodeobs,holoobs09,intholo,holoissue,healworld}.
By far, various observational constraints on this model all indicate
that the parameter $c$ is less than 1, implying that the holographic
dark energy would lead to a phantom universe with big rip as its
ultimate fate \cite{holodeobs,holoobs09}. Nevertheless, the
appearance of the cosmic doomsday will definitely ruin the
theoretical foundation of the holographic dark energy model, namely,
the effective quantum field theory; see Ref.~\cite{healworld} for
detailed discussions. To evade this difficulty, one may consider the
extra dimension effects in this model \cite{healworld}: The high
energy correction from the extra dimensions could erase the big-rip
singularity and leads to a de Sitter finale for the holographic
cosmos. Another way of avoiding the big rip is to consider some
phenomenological interaction between holographic dark energy and
matter \cite{intholo,holoissue}. With the help of the interaction,
the big rip might be avoided due to the occurrence of an attractor
solution in which the effective equations of state of dark energy
and matter become identical in the far future. However, it should be
pointed out that in order to avoid the phantom-like universe the
phenomenological interaction term should be tuned to satisfy some
specific condition, i.e., the interaction must be strong enough; for
detailed discussions see Ref.~\cite{holoissue}. The only way of
acquiring the knowledge of the interaction strength is from the
observational data fitting. In this paper, we will explore the
possible phenomenological interactions between holographic dark
energy and matter by using the latest observational data.

Another important mission of this paper is to probe the spatial
curvature of the universe in the holographic dark energy model.
Actually, the current observational data are not accurate enough to
distinguish between the dynamical effects of dark energy and spatial
curvature of the universe, owing to the degeneracy between them.
Indeed, in fitting the equation of state of dark energy, $w(z)$, the
inclusion of $\Omega_{k0}$ as an additional parameter would dilute
constraints on $w(z)$. While most inflation models in which the
inflationary periods last for much longer than 60 $e$-folds predict
a spatially flat universe, $\Omega_{k0}\sim 10^{-5}$, the current
constraint on $\Omega_{k0}$ is three orders of magnitude larger than
this inflation prediction. As argued in Ref.~\cite{Clarkson:2007bc},
the studies of dark energy, and in particular, of observational
data, should include $\Omega_{k0}$ as a parameter to be fitted
alongside the $w(z)$ parameters. So, in this paper, we will
constrain $\Omega_{k0}$ in the holographic dark energy model in
light of the latest observational data. Of course, we will also let
the interaction and spatial curvature simultaneously be free
parameters, and explore the possible degeneracy between them.

\section{Holographic dark energy with interaction and spatial curvature}
In this section we shall describe the interacting holographic dark
energy in a non-flat universe. In a spatially non-flat
Friedmann-Robertson-Walker universe, the Friedmann equation can be
written as
\begin{equation}
\label{eq:FE1} 3M_{Pl}^2\left(H^2+{k \over a^2}\right)=\rho,
\end{equation}
where $\rho=\rho_\Lambda+\rho_m$. We define
\begin{equation}
\label{eq:DefOmega_k} \Omega_k={k \over
H^2a^2}=\Omega_{k0}\Big({H_0\over aH}\Big)^2,\ \ \ \ \
\Omega_{\Lambda}={\rho_{\Lambda} \over \rho_c},\ \ \ \ \
\Omega_{m}={\rho_{m} \over \rho_c},
\end{equation}
where $\rho_c=3M_{Pl}^2H^2$ is the critical density of the universe,
thus we have
\begin{equation}
\label{eq:AllOmega} 1+\Omega_k=\Omega_m+\Omega_{\Lambda}.
\end{equation}

Consider, now, some interaction between holographic dark energy and
matter:
\begin{equation}
\label{eq:IHDEeq0}\dot\rho_m+3H\rho_m=Q,
\end{equation}
\begin{equation}
\label{eq:IHDEeq1}\dot\rho_{\Lambda}+3H(\rho_{\Lambda}+p_{\Lambda})=-Q,
\end{equation}
where $Q$ denotes the phenomenological interaction term. Owing to
the lack of the knowledge of micro-origin of the interaction, we
simply follow other work on the interacting holographic dark energy
and parameterize the interaction term generally as
$Q=-3H(b_1\rho_\Lambda+b_2\rho_m)$, where $b_1$ and $b_2$ are the
coupling constants. For reducing the complication and the number of
parameters, furthermore, we only consider the following three cases:
(a) $b_1=b$ and $b_2=0$, (b) $b_1=b_2=b$, (c) $b_1=0$ and $b_2=b$.
We denote the interaction term $Q$ in these three cases as
\begin{eqnarray}
\label{eq:I1}&&Q_1=-3bH\rho_{\Lambda},\\
\label{eq:I2}&&Q_2=-3bH(\rho_{\Lambda}+\rho_m),\\
\label{eq:I3}&&Q_3=-3bH\rho_m.
\end{eqnarray}
For convenience, in the following we uniformly express the
interaction term as $Q_i=-3bH\rho_c\Omega_{i}$, where
$\Omega_{i}=\Omega_{\Lambda}$, 1 and $\Omega_m$, for $i=1$, $2$ and
$3$, respectively. Note that according to our convention $b<0$ means
that dark energy decays to matter. Moreover, it should be pointed
out that $b>0$ will lead to unphysical consequences in physics,
since $\rho_m$ will become negative and $\Omega_\Lambda$ will be
greater than 1 in the far future. So, the parameter $b$ is always
assumed to be negative in the literature. However, in the present
work, instead of making such an assumption on $b$, we let $b$ be
totally free and let the observational data tell us the true story
about the holographic dark energy, no matter whether the ultimate
fate of the universe is ridiculous or not.

From the definition of holographic dark energy (\ref{eq:rhode0}), we
have
\begin{equation}
\label{eq:Omegade}\Omega_{\Lambda}={c^2\over H^2L^2},
\end{equation}
or equivalently,
\begin{equation}
\label{ea:L0} L={c \over H\sqrt{\Omega_{\Lambda}}}.
\end{equation}
Thus, we easily get
\begin{equation}
\label{eq:dotL}\dot L={d\over dt}\left({c \over
H\sqrt{\Omega_{\Lambda}}}\right) =\left(-{\dot H \over
H}-{1\over2}{\dot{\Omega_{\Lambda}}\over \Omega_{\Lambda}} \right){c
\over H \sqrt{\Omega_{\Lambda}}}.
\end{equation}

Following Ref.~\cite{holode1}, in a non-flat universe the IR cut-off
length scale $L$ takes the form
\begin{equation}
\label{eq:L1} L=ar(t),
\end{equation}
and $r(t)$ satisfies
\begin{equation}
\label{eq:r(t)1}  \int_0^{r(t)} {dr \over
\sqrt{1-kr^2}}=\int_t^{+\infty}{dt\over a(t)}.
\end{equation}
Thus, we have
\begin{equation}
\label{eq:r(t)2} r(t)={1\over\sqrt{k}}\sin
\Big(\sqrt{k}\int_t^{+\infty} {dt \over a}\Big)={1\over\sqrt{k}}\sin
\Big(\sqrt{k}\int_{a(t)}^{+\infty} {da \over {Ha^2}}\Big).
\end{equation}
Equation (\ref{eq:L1}) leads to another equation about $r(t)$,
namely,
\begin{equation}
\label{eq:r(t)3.0} r(t)={L\over
a}={c\over\sqrt{\Omega_{\Lambda}}Ha}.
\end{equation}
Combining Eqs. (\ref{eq:r(t)2}) and (\ref{eq:r(t)3.0}) yields
\begin{equation}
\label{eq:OL1.1}\sqrt{k}\int_t^{+\infty}{dt\over
a}=\arcsin{c\sqrt{k}\over \sqrt{\Omega_{\Lambda}}aH}.
\end{equation}
Taking derivative of Eq.~(\ref{eq:OL1.1}) with respect to $t$, one
can get
\begin{equation}
\label{eq:OL1.3}\sqrt{{\Omega_{\Lambda}H^2\over c^2}-{k\over
a^2}}={\dot\Omega_{\Lambda}\over2\Omega_{\Lambda}}+H+{\dot H\over
H}.
\end{equation}

Let us plus Eqs.~(\ref{eq:IHDEeq0}) and (\ref{eq:IHDEeq1}) together,
and then we can obtain the form of $p_{\Lambda}$:
\begin{eqnarray}
&&{(\dot{\rho}_{c}+\dot{\rho}_k)}+3H(\rho_{\Lambda}+\rho_m+p_{\Lambda})=0 \nonumber\\
&&\Rightarrow
{(\dot{\rho}_{c}+\dot{\rho}_k)}+3H(\rho_{c}+\rho_k+p_{\Lambda})=0 \nonumber\\
&&\Rightarrow p_{\Lambda}=-{1\over3H}\Big(2{\dot H\over
H}\rho_c-2{\dot a\over a}\rho_k \Big)-\rho_c-\rho_k .
\end{eqnarray}
Substituting $p_{\Lambda}$ into Eq. (\ref{eq:IHDEeq1}), we have
\begin{equation}
\Big(2{\dot H\over H}+{\dot\Omega_{\Lambda}\over\Omega_{\Lambda}}+3H
\Big)\rho_{\Lambda}-H\rho_k-\Big(2{\dot H\over H}+3H\Big)\rho_c
=3bH\rho_i,
\end{equation}
where $\rho_i$ (with $i=1\sim 3$) denotes $\rho_1=\rho_{\Lambda}$,
$\rho_2=\rho_{\Lambda}+\rho_{m}=\rho_c+\rho_k$, and
$\rho_3=\rho_m=\rho_c+\rho_k-\rho_{\Lambda}$, respectively. Divided
the above equation by $\rho_c$, we get an equation containing $\dot
H$ and $\dot \Omega_{\Lambda}$,
\begin{equation}
\label{eq:OH2} 2(\Omega_{\Lambda}-1){\dot H\over
H}+\dot\Omega_{\Lambda}+H(3\Omega_{\Lambda}-3-\Omega_k)=3bH\Omega_i,
\end{equation}
where $\Omega_i=\Omega_{\Lambda}$, $\Omega_{\Lambda}+\Omega_k$, and
$1+\Omega_k-\Omega_{\Lambda}$, for $i=1$, 2, and 3, respectively.
Combining this equation with Eq.~(\ref{eq:OL1.3}), we eventually
obtain the following two equations governing the dynamical evolution
of the interacting holographic dark energy in a non-flat universe,
\begin{equation}
\label{eq:OH3}{1\over H/H_0}{d\over dz}\left({H\over H_0}\right)
=-{\Omega_{\Lambda}\over
1+z}\left({3\Omega_{\Lambda}-\Omega_k-3-3b\Omega_i\over2\Omega_{\Lambda}}-1+\sqrt{{\Omega_{\Lambda}\over
c^2}-{\Omega_{k0}(1+z)^2\over(H/H_0)^2}} \right),
\end{equation}
\begin{equation}
\label{eq:OH4}{d\Omega_{\Lambda}\over
dz}=-{2\Omega_{\Lambda}(1-\Omega_{\Lambda})\over
1+z}\left(\sqrt{{\Omega_{\Lambda}\over
c^2}-{\Omega_{k0}(1+z)^2\over(H/H_0)^2}}-1-{3\Omega_{\Lambda}-{(1+z)^2\Omega_{k0}\over(H/
H_0)^2}-3-3b\Omega_i\over 2(1-\Omega_{\Lambda})}\right).
\end{equation}
These two equations can be solved numerically and will be used in
the data analysis procedure.

\section{Observational Constraints}

In the holographic dark energy model with interaction and spatial
curvature, there are four free parameters: $c$, $\Omega_{m0}$, $b$,
and $\Omega_{k0}$. In this section, we shall constrain these
parameters of the holographic dark energy model by using the latest
observational data. For decreasing the complication, let us close
some parameters in turn. We shall consider the following four cases:
(a) the model of holographic dark energy without interaction and
spatial curvature (namely, $b=0$ and $\Omega_{k0}=0$), denoted as
HDE; (b) the model of holographic dark energy with spatial curvature
but without interaction (namely, $\Omega_{k0}\neq 0$ but $b=0$),
denoted as KHDE; (c) the model of holographic dark energy with
interaction but without spatial curvature (namely, $b\neq 0$ but
$\Omega_{k0}=0$), denoted as IHDE; (d) the model of holographic dark
energy with interaction and spatial curvature (namely, $b\neq 0$ and
$\Omega_{k0}\neq 0$), denoted as KIHDE.

\subsection{Observational data used}

The observational data we use in this paper include the Constitution
SNIa sample, the shift parameter of the CMB given by the five-year
Wilkinson Microwave Anisotropy Probe (WMAP5) observations, and the
baryon acoustic oscillation (BAO) measurement from the Sloan Digital
Sky Survey (SDSS).

\subsubsection{Type Ia supernovae}

For the SNIa data, we use the latest Constitution sample including
397 data that are given in terms of the distance modulus $\mu_{
obs}(z_i)$ compiled in Table 1 of Ref.~\cite{Hicken2}. The
theoretical distance modulus is defined as
\begin{equation}
\mu_{th}(z_i)\equiv 5 \log_{10} {D_L(z_i)} +\mu_0,
\end{equation}
where $\mu_0\equiv 42.38-5\log_{10}h$ with $h$ the Hubble constant
$H_0$ in units of 100 km/s/Mpc, and the Hubble-free luminosity
distance $D_L=H_0d_L$ is
\begin{equation}
D_L(z)={1+z\over \sqrt{|\Omega_{k0}|}}\textrm{sinn}\Big(
\sqrt{|\Omega_{k0}|}\int_0^z{dz'\over E(z')} \Big),
\end{equation}
where $E(z)\equiv H(z)/H_0$, and
\begin{displaymath}
{\textrm{sinn}\left(\sqrt{|\Omega_{k0}|}x\right)\over
\sqrt{|\Omega_{k0}|}} = \left\{
\begin{array}{ll}
{\textrm{sin}(\sqrt{|\Omega_{k0}|}x)/
\sqrt{|\Omega_{k0}|}}, & \textrm{if $\Omega_{k0}>0$},\\
x, & \textrm{if $\Omega_{k0}=0$},\\
{\textrm{sinh}(\sqrt{|\Omega_{k0}|}x)/
 \sqrt{|\Omega_{k0}|}}, &
\textrm{if $\Omega_{k0}<0$}.
\end{array} \right.
\end{displaymath}
The $\chi^2$ for the SNIa data is
\begin{equation}
\chi^2_{SN}=\sum\limits_{i=1}^{397}{[\mu_{obs}(z_i)-\mu_{th}(z_i)]^2\over
\sigma_i^2},\label{ochisn}
\end{equation}
where $\mu_{obs}(z_i)$ and $\sigma_i$ are the observed value and the
corresponding 1$\sigma$ error of distance modulus for each
supernova, respectively.

\subsubsection{Baryon acoustic oscillation}

For the BAO data, we consider the parameter $A$ from the measurement
of the BAO peak in the distribution of SDSS luminous red galaxies,
which is defined as \cite{Eisenstein}
\begin{equation}
A\equiv{\sqrt{\Omega_{m0}}\over {E(z_{b})}^{1\over3}}\left[{1\over
z_{b}\sqrt{|\Omega_{k0}|}}\textrm{sinn}\Big(\sqrt{|\Omega_{k}|}\int_0^{z_{b}}{dz'\over
E(z')} \Big)\right]^{2\over3},
\end{equation}
where $z_{b}=0.35$. The SDSS BAO measurement \cite{Eisenstein} gives
$A_{obs}=0.469\,(n_s/0.98)^{-0.35}\pm 0.017$, where the scalar
spectral index is taken to be $n_s=0.960$ as measured by WMAP5
\cite{Komatsu}. It is widely believed that $A$ is nearly model
independent and can provide robust constraint as complement to SNIa
data. The $\chi^2$ for the BAO data is
\begin{equation}\label{chiLSS}
\chi^2_{BAO}=\frac{(A-A_{obs})^2}{\sigma_A^2},
\end{equation}
where the corresponding $1\sigma$ error is $\sigma_A=0.017$.

\subsubsection{Cosmic microwave background}

For the CMB data, we use the CMB shift parameter $R$ given by
\cite{Bond,ywang3}
\begin{equation}
R\equiv{\sqrt{\Omega_{m0}}\over
\sqrt{|\Omega_{k0}|}}\textrm{sinn}\Big(\sqrt{|\Omega_{k0}|}\int_0^{z_{rec}}{dz'\over
E(z')} \Big),
\end{equation}
where the redshift of recombination $z_{rec}=1090$ \cite{Komatsu}.
The shift parameter $R$ relates the angular diameter distance to the
last scattering surface, the comoving size of the sound horizon at
$z_{rec}$ and the angular scale of the first acoustic peak in CMB
power spectrum of temperature fluctuations \cite{Bond,ywang3}. The
current measured value of $R$ is $R_{obs}=1.710\pm 0.019$ from WMAP5
\cite{Komatsu}. It should be noted that, different from the SNIa and
BAO data, the $R$ parameter can provide the information about the
universe at very high redshift. The $\chi^2$ for the CMB data is
\begin{equation}\label{chiCMB}
\chi^2_{CMB}=\frac{(R-R_{obs})^2}{\sigma_R^2},
\end{equation}
where the corresponding $1\sigma$ error is $\sigma_R=0.019$.

\begin{table} \caption{Fit results for the holographic dark energy model. }
\begin{center}
\label{table1}
\begin{tabular}{ccccccc}
  \hline\hline
Model   &           $\Omega_{m0}$      &             $c$                 &               $\Omega_{k0}$                 &                    $ b $                    &     $\chi^2_{min}$    &     $\Delta\chi^2_{min}$    \\
  \hline
  HDE    ~&~~~$0.277^{+0.022}_{-0.021}$ ~~&~~ $0.818^{+0.113}_{-0.097}$~~ &                                             &                                           & ~~~   465.912    ~~~  & ~~~       0        ~~~  \\
  \hline
  KHDE   ~&~~~$0.278^{+0.037}_{-0.035}$ ~~&~~ $0.815^{+0.179}_{-0.139}$~~ & ~~ $(7.7\times10^{-4})^{+0.018}_{-0.019}$ ~~&~~                                         & ~~~   465.906    ~~~  & ~~~       0.006    ~~~  \\
  \hline
  IHDE1   &~~ $0.277^{+0.035}_{-0.034}$ ~~&~~ $0.818^{+0.197}_{-0.257}$~~ & ~~                                        ~~&~~  $(6.1\times10^{-5})^{+0.036}_{-0.025}$ & ~~~   465.911    ~~~  & ~~~       0.001    ~~~  \\
  \hline
  IHDE2   &~~ $0.277^{+0.034}_{-0.036}$ ~~&~~ $0.816^{+0.170}_{-0.223}$~~ & ~~                                        ~~&~~  $(1.6\times10^{-4})^{+0.009}_{-0.008}$ & ~~~   465.910    ~~~  & ~~~       0.002    ~~~  \\
  \hline
  IHDE3   &~~ $0.277^{+0.034}_{-0.036}$ ~~&~~ $0.815^{+0.164}_{-0.209}$~~ & ~~                                        ~~&~~  $(3.0\times10^{-4})^{+0.011}_{-0.011}$ & ~~~   465.909    ~~~  & ~~~       0.003    ~~~  \\
  \hline
  KIHDE1  &~~ $0.281^{+0.047}_{-0.043}$ ~~&~~ $0.977^{+0.563}_{-0.551}$~~ & ~~          $0.030^{+0.066}_{-0.127}$     ~~&~~ $       -0.046^{+0.243}_{-0.102}$       & ~~~   465.697    ~~~  & ~~~       0.315    ~~~  \\
  \hline
  KIHDE2  &~~ $0.281^{+0.047}_{-0.044}$ ~~&~~ $0.974^{+0.559}_{-0.475}$~~ & ~~          $0.030^{+0.070}_{-0.100}$     ~~&~~ $       -0.042^{+0.191}_{-0.073}$       & ~~~   465.700    ~~~  & ~~~       0.312    ~~~  \\
  \hline
  KIHDE3  &~~ $0.280^{+0.045}_{-0.042}$ ~~&~~ $0.961^{+0.231}_{-0.499}$~~ & ~~          $0.061^{+0.038}_{-0.210}$     ~~&~~ $       -0.048^{+0.113}_{-0.042}$       & ~~~   465.719    ~~~  & ~~~       0.293    ~~~  \\
  \hline\hline
\end{tabular}
\end{center}
\end{table}

\begin{figure}
\includegraphics[width=0.35\textwidth]{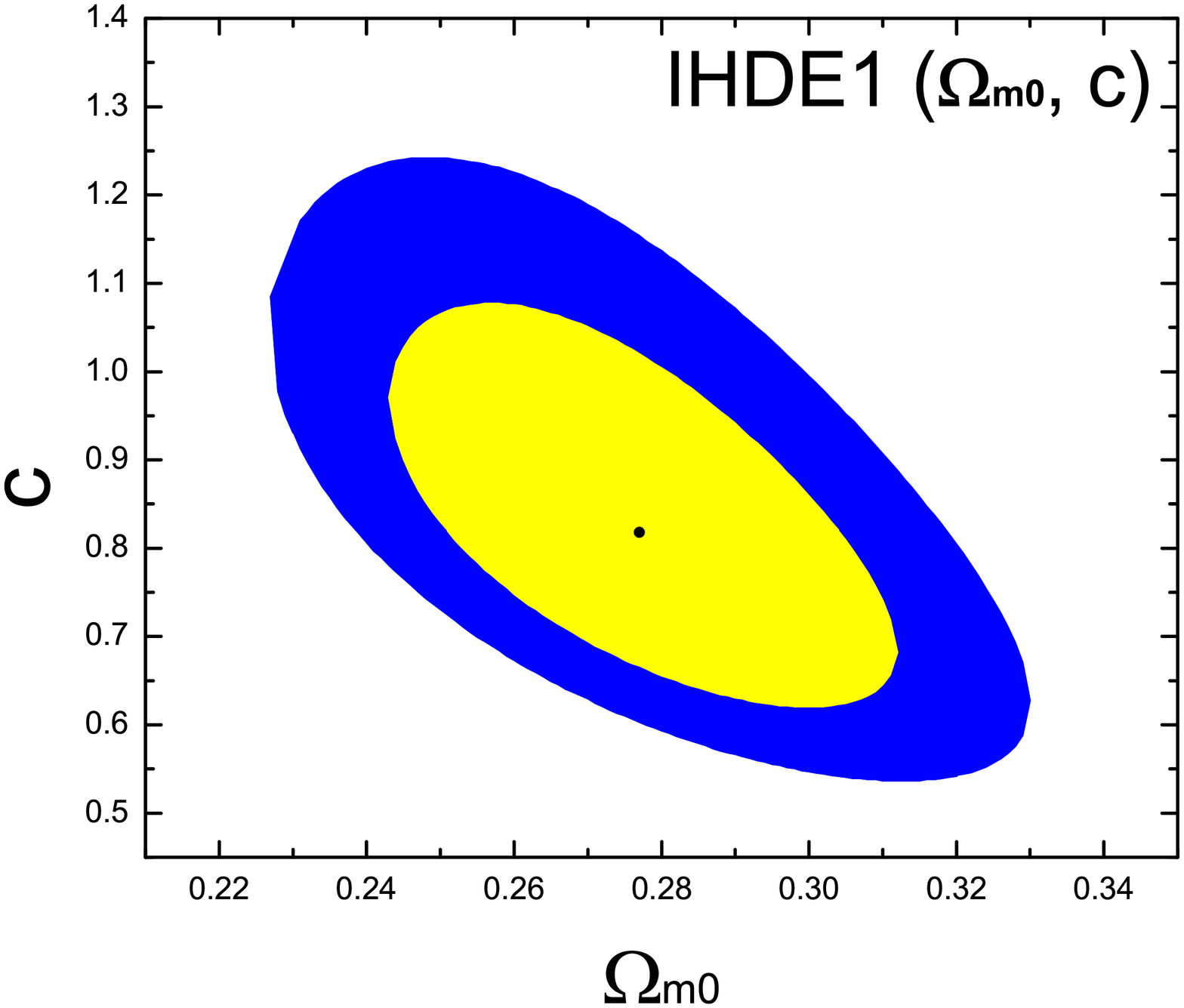}\hskip.5cm
\includegraphics[width=0.35\textwidth]{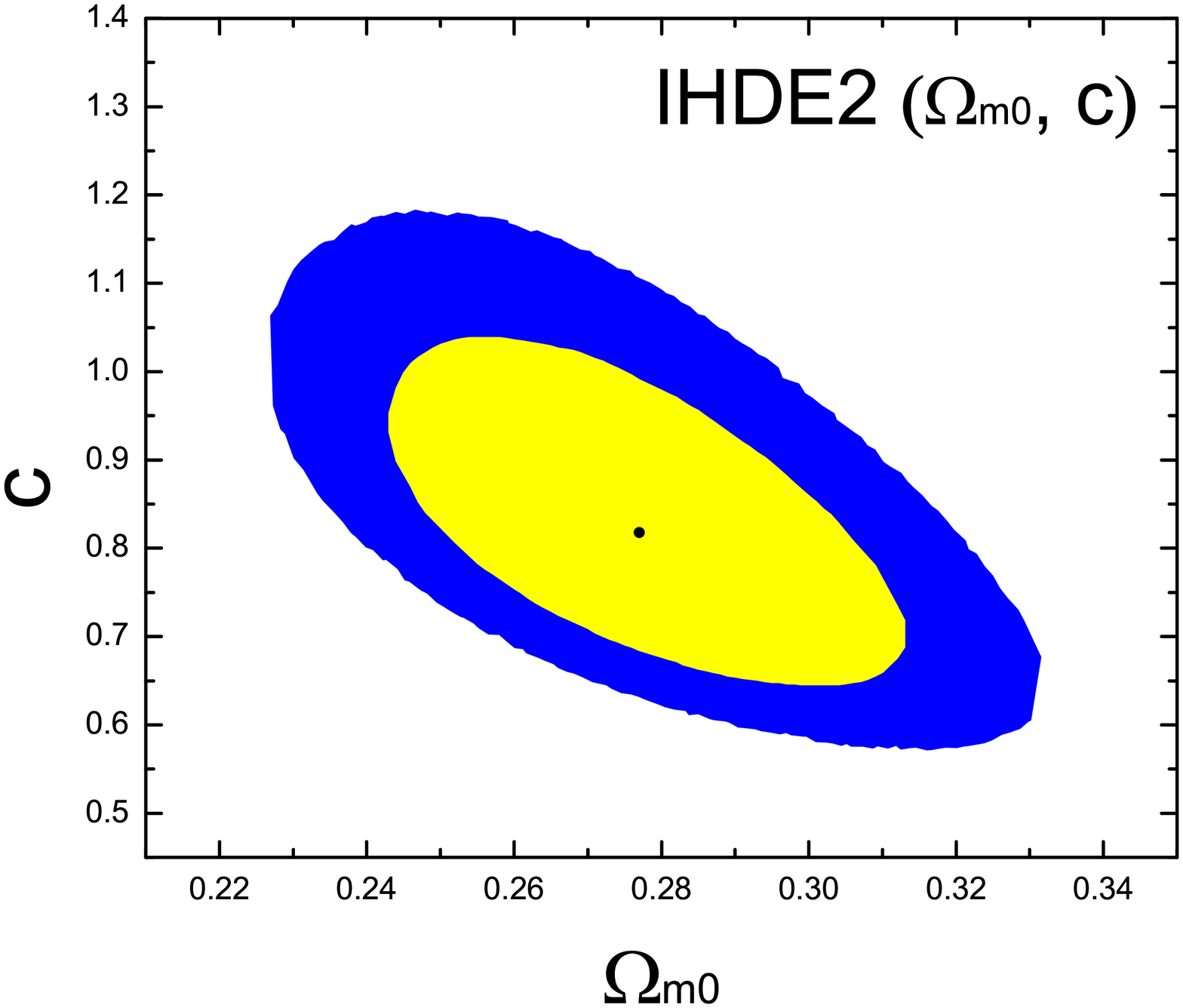}\vskip.5cm
\includegraphics[width=0.35\textwidth]{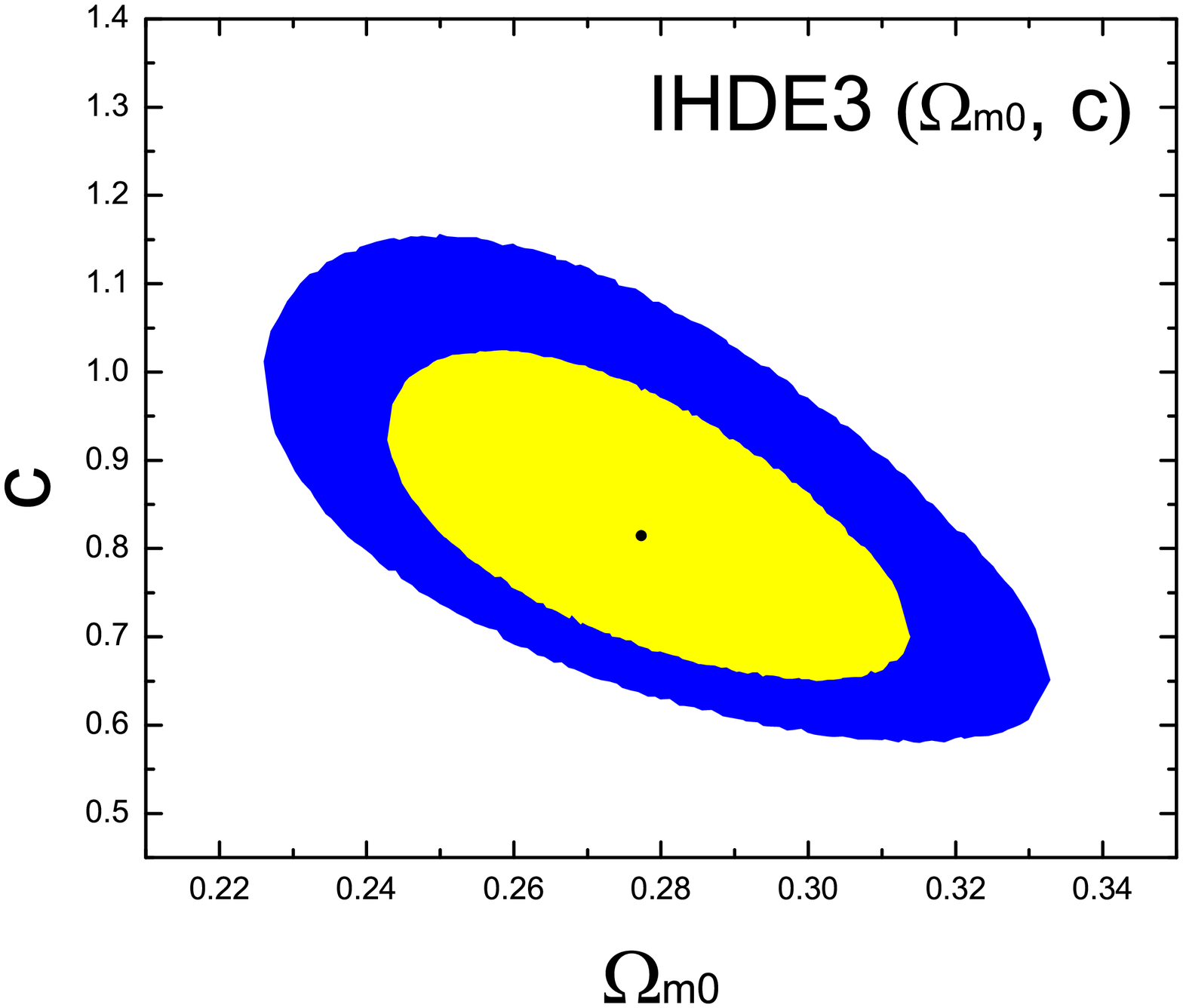}\hskip.5cm
\includegraphics[width=0.35\textwidth]{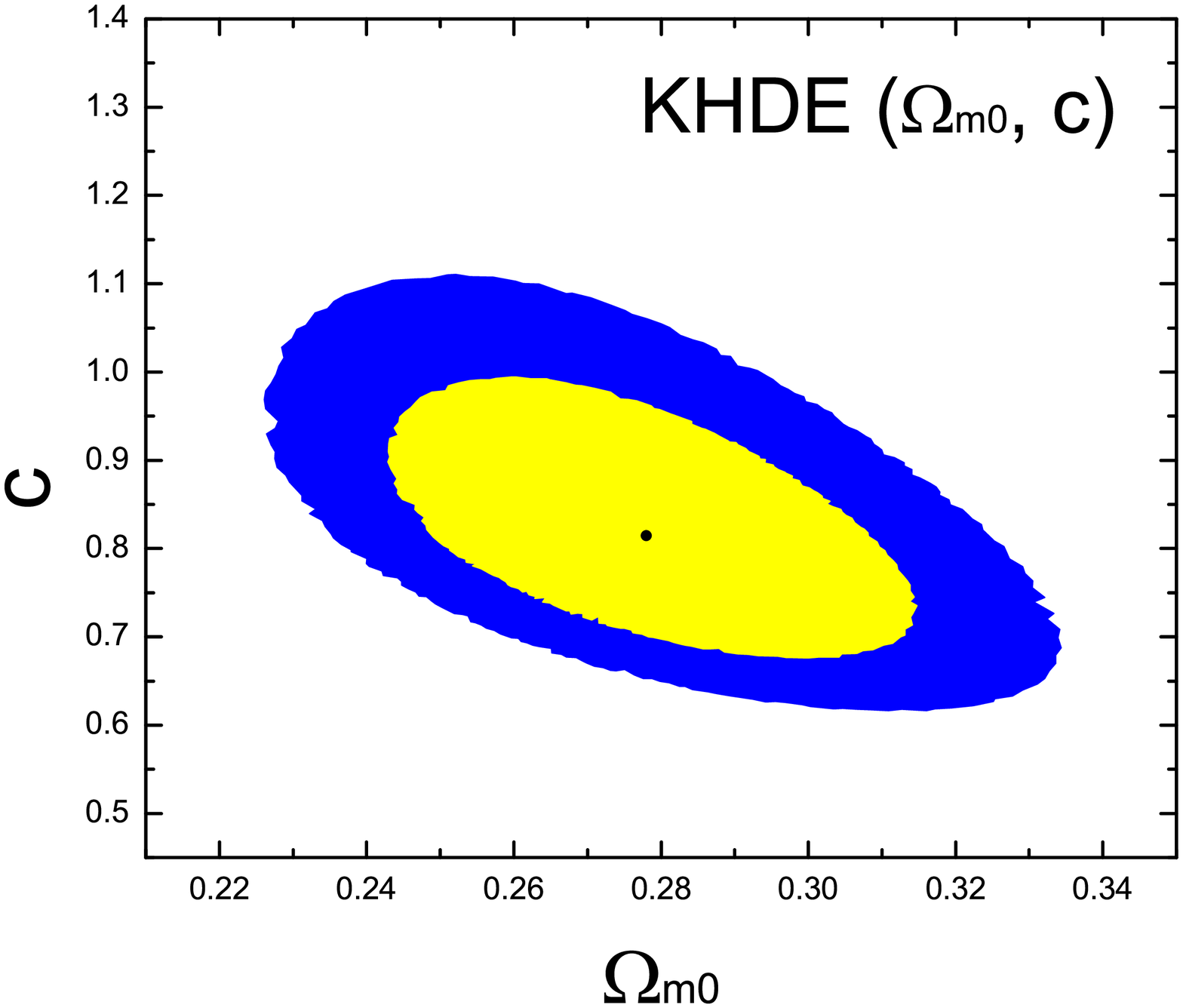}
\caption{Probability contours at $68.3\%$ and $95.4\%$ confidence
levels in the $\Omega_{m0}-c$ plane, for the three IHDE models and
the KHDE model.}\label{fig-Omega_m0_and_c}
\end{figure}

\begin{figure}
\includegraphics[width=0.35\textwidth]{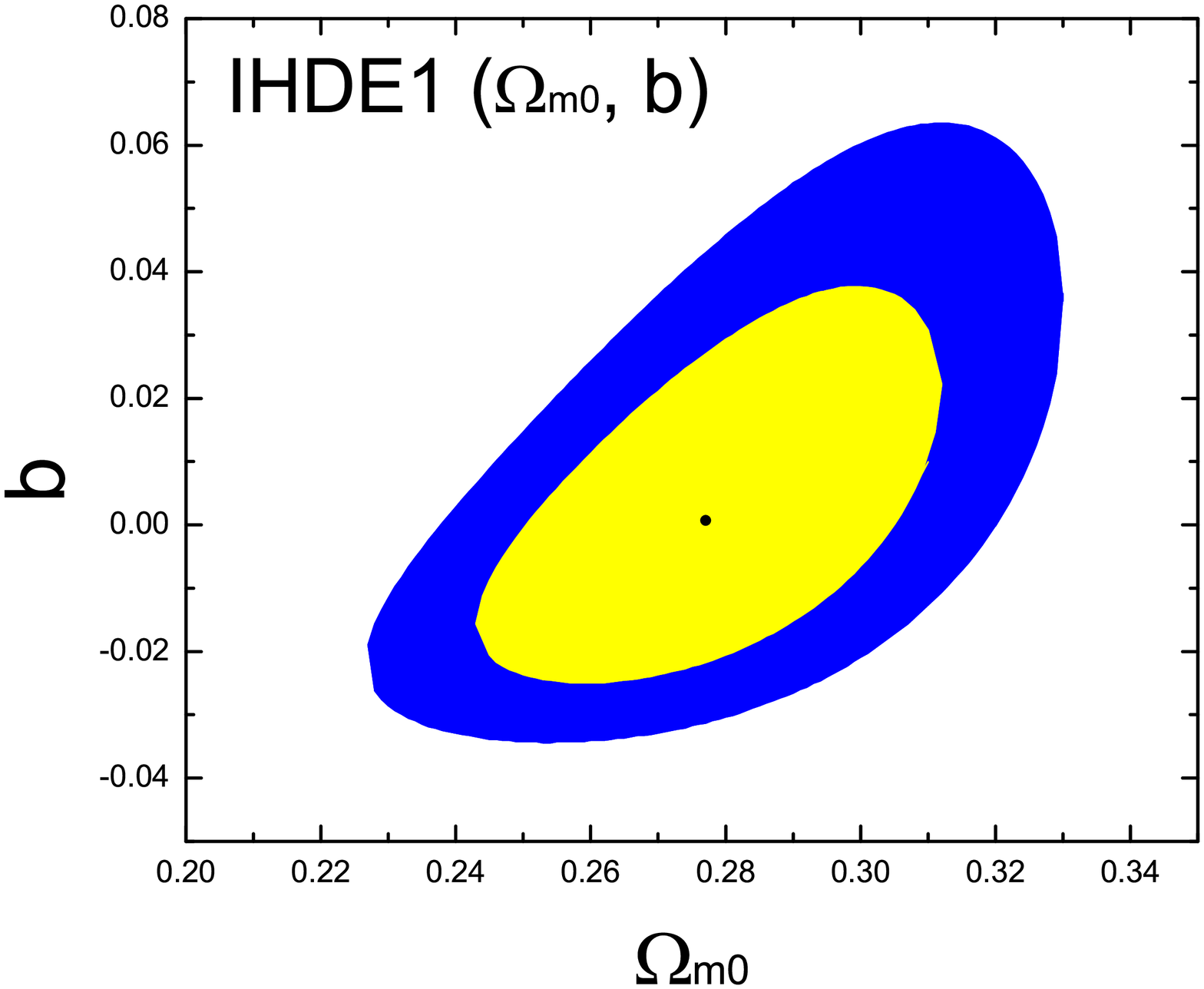}\hskip.5cm
\includegraphics[width=0.35\textwidth]{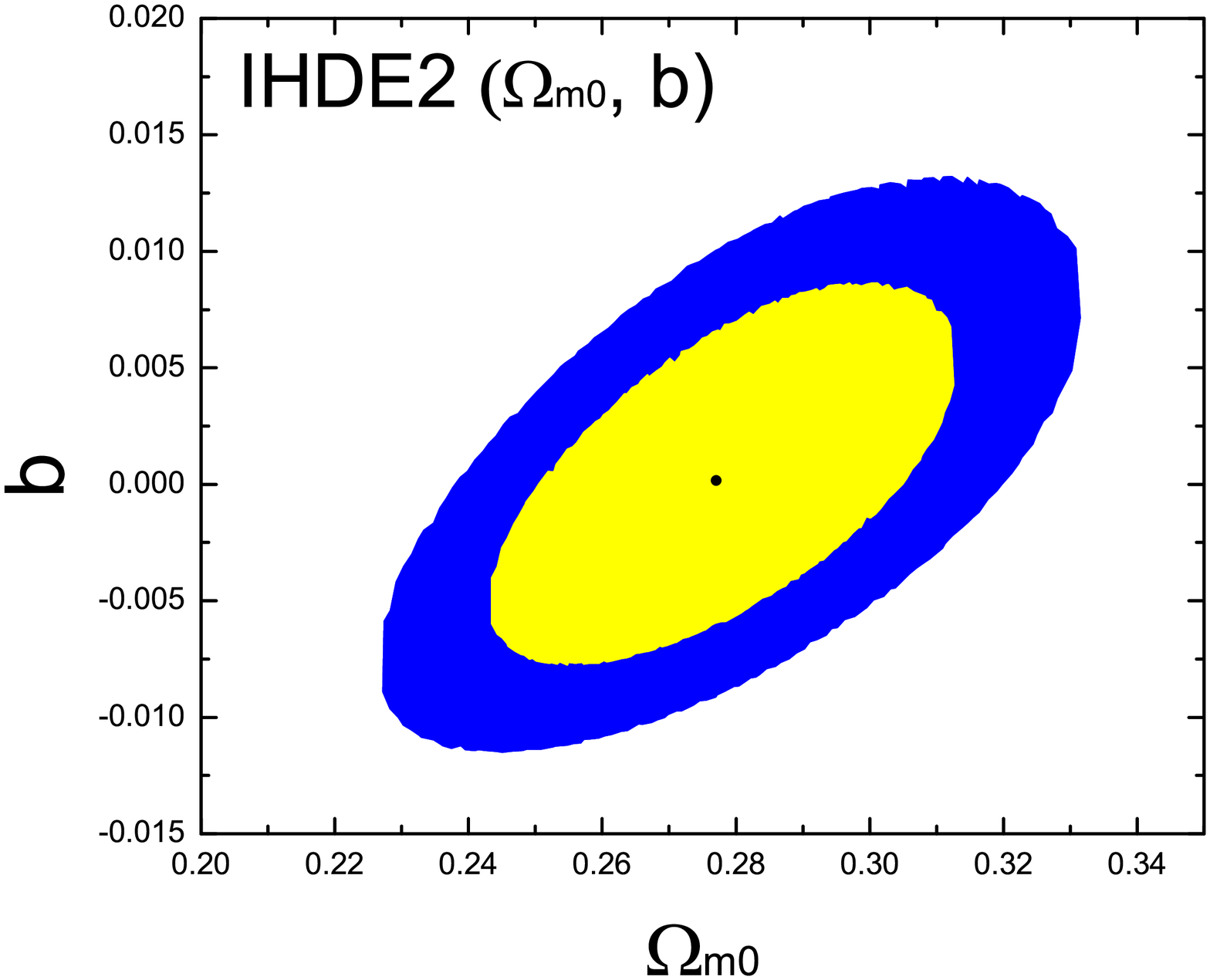}\vskip.5cm
\includegraphics[width=0.35\textwidth]{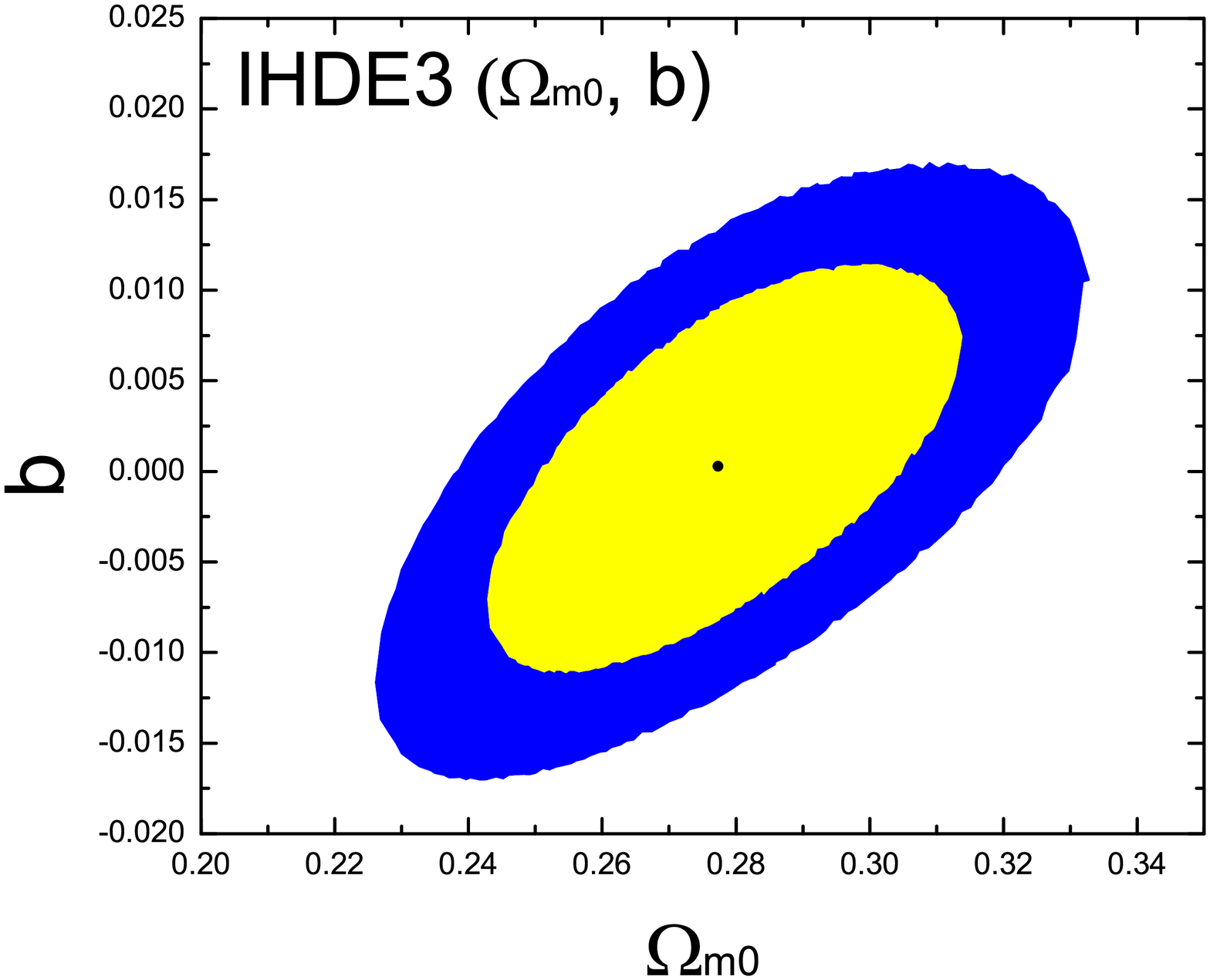}
\caption{Probability contours at $68.3\%$ and $95.4\%$ confidence
levels in the $\Omega_{m0}-b$ plane, for the three IHDE
models.}\label{fig-Omega_m0_and_b}
\end{figure}

\begin{figure}
\includegraphics[width=0.35\textwidth]{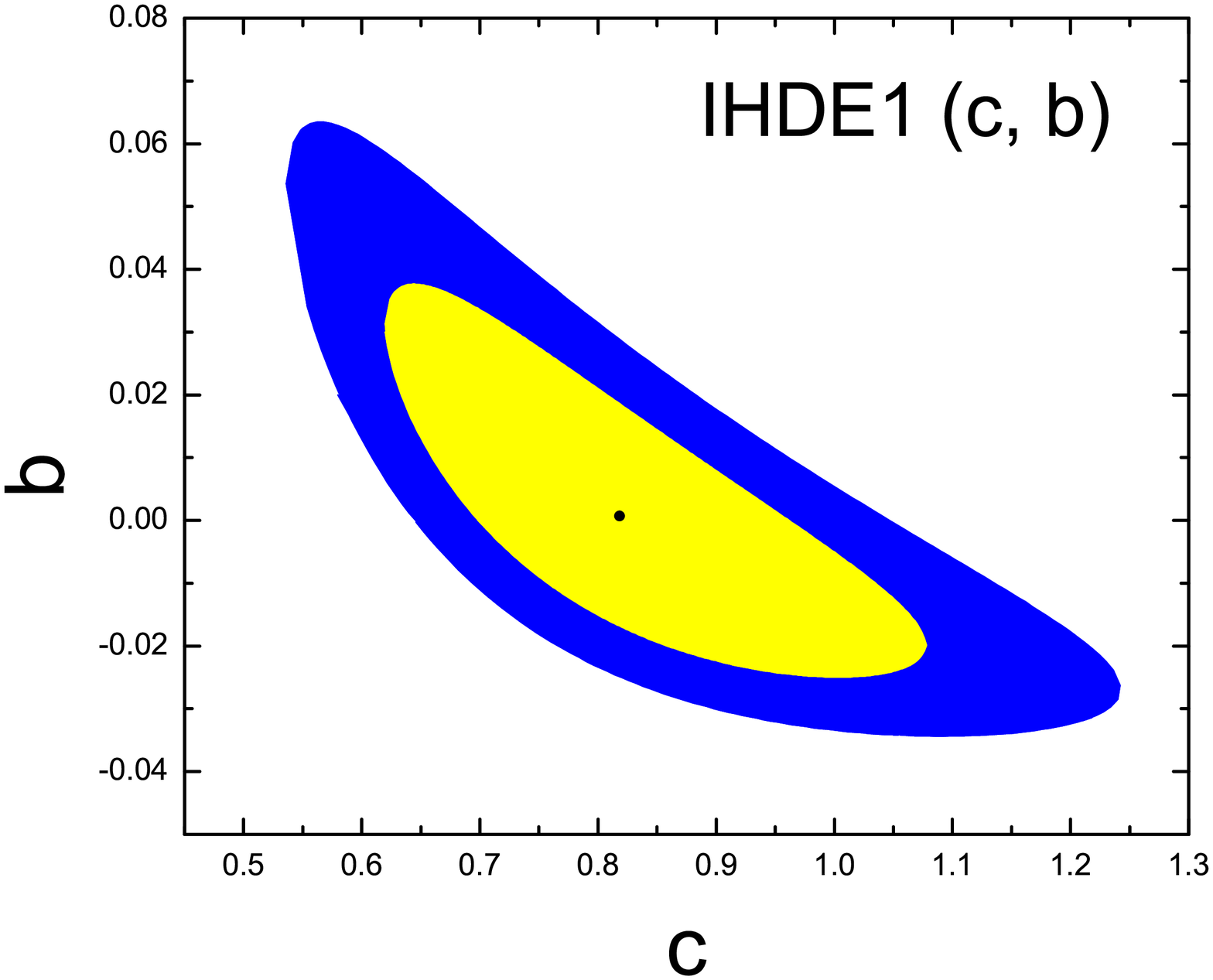}\hskip.5cm
\includegraphics[width=0.35\textwidth]{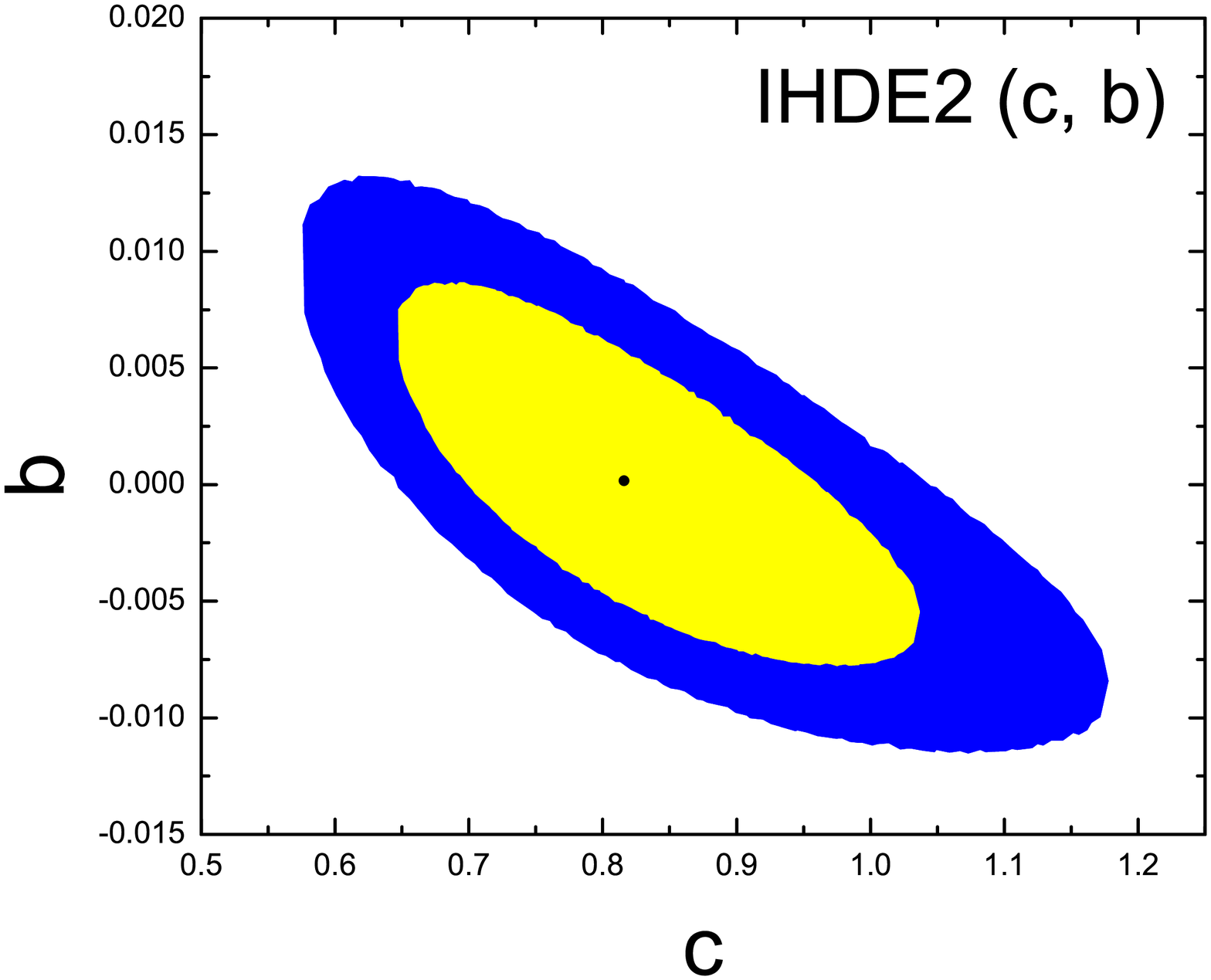}\vskip.5cm
\includegraphics[width=0.35\textwidth]{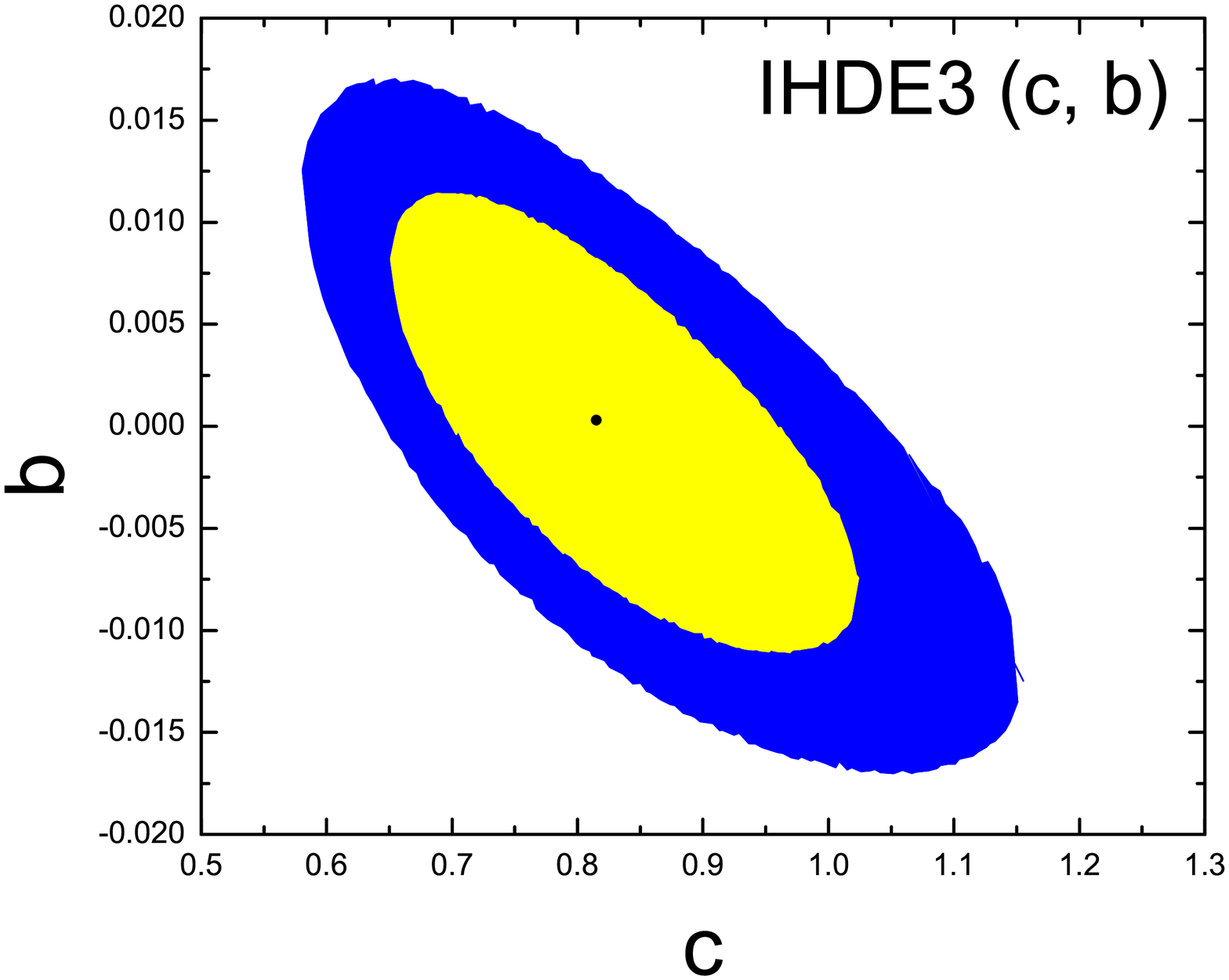}
\caption{Probability contours at $68.3\%$ and $95.4\%$ confidence
levels in the $c-b$ plane, for the three IHDE
models.}\label{fig-c_and_b}
\end{figure}

\begin{figure}
\includegraphics[width=0.35\textwidth]{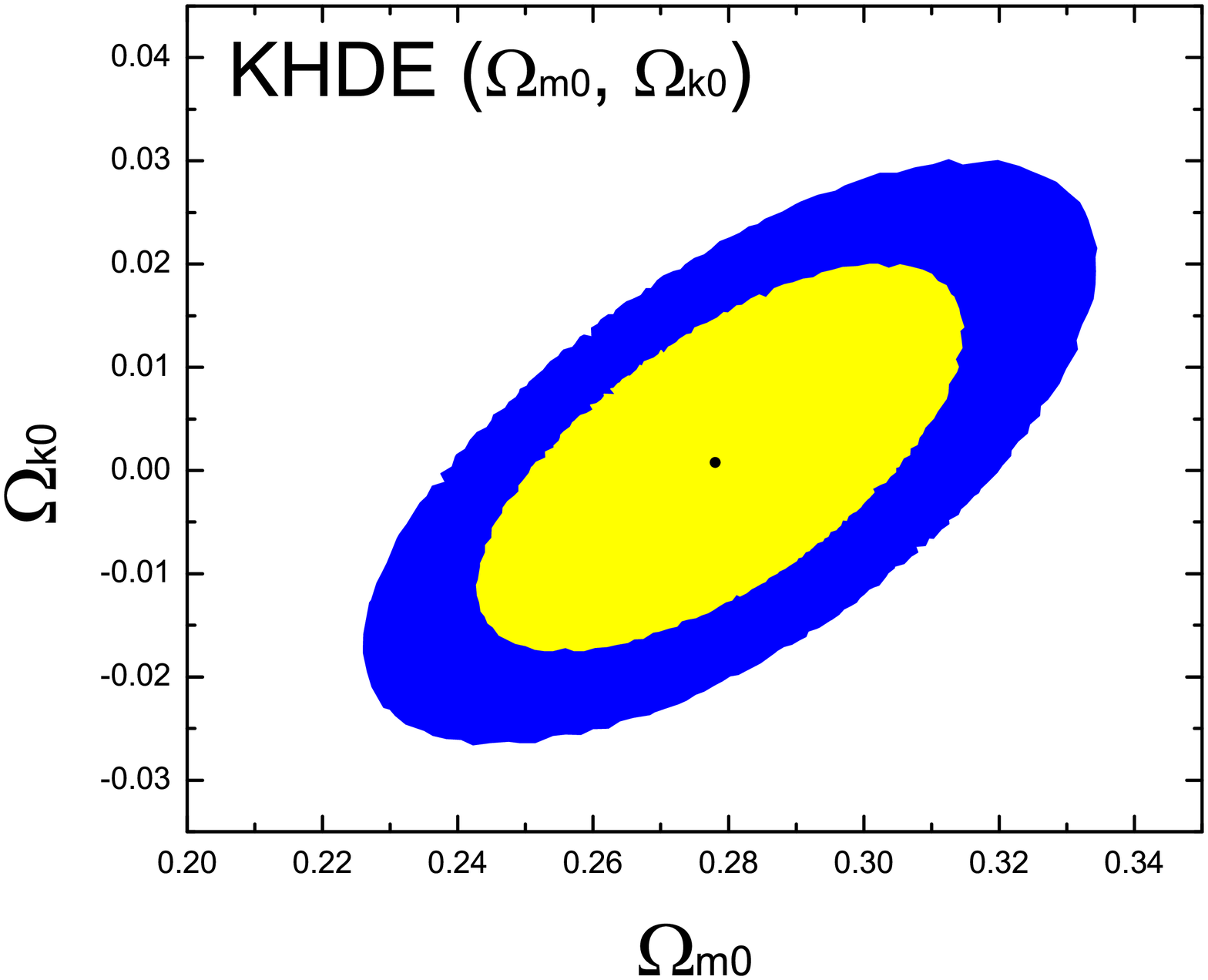}\hskip.5cm
\includegraphics[width=0.35\textwidth]{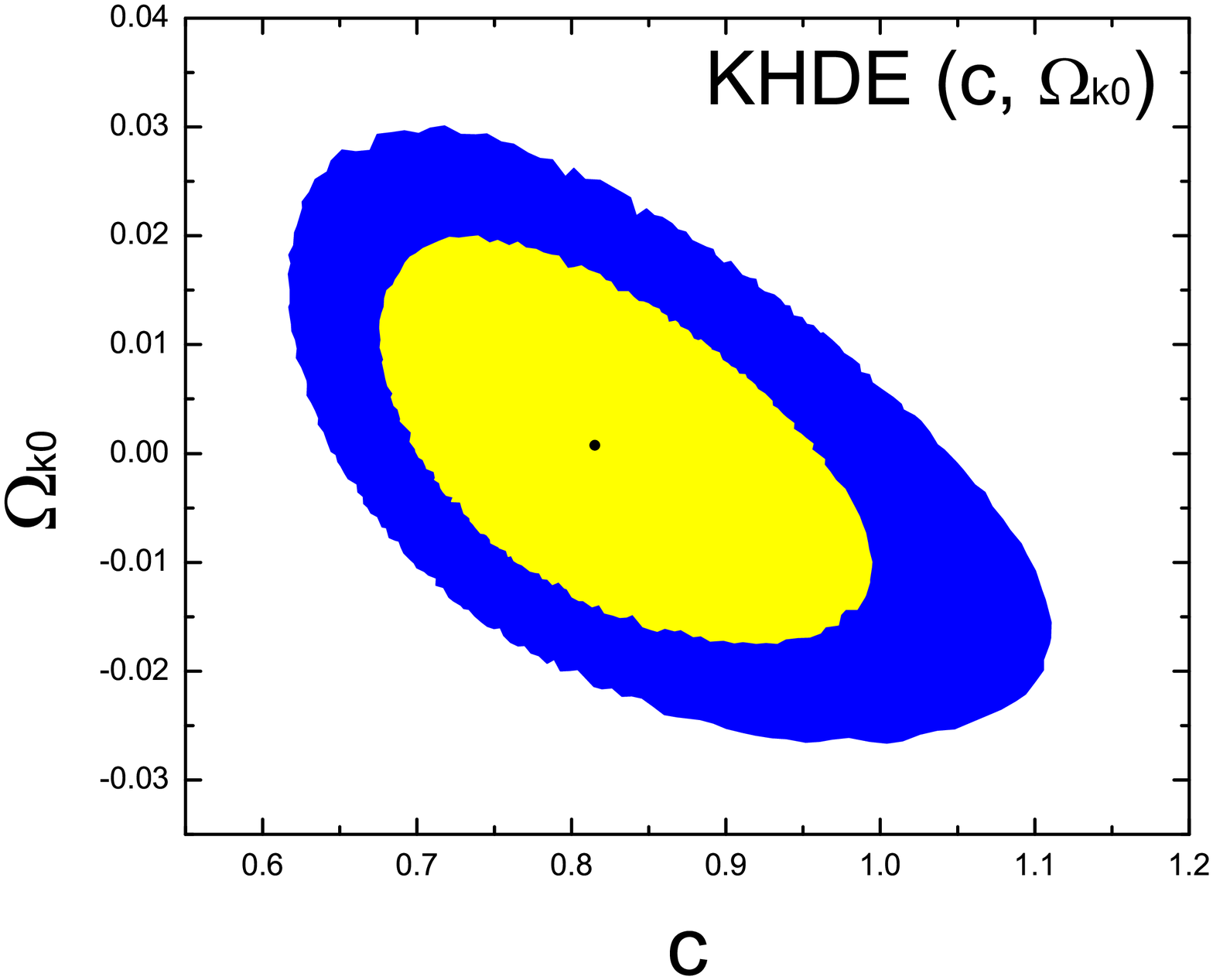}
\caption{Probability contours at $68.3\%$ and $95.4\%$ confidence
levels in the $\Omega_{m0}-\Omega_{k0}$ and $c-\Omega_{k0}$ planes,
for the KHDE model.}\label{KHDE}
\end{figure}


\begin{figure}
\includegraphics[width=0.35\textwidth]{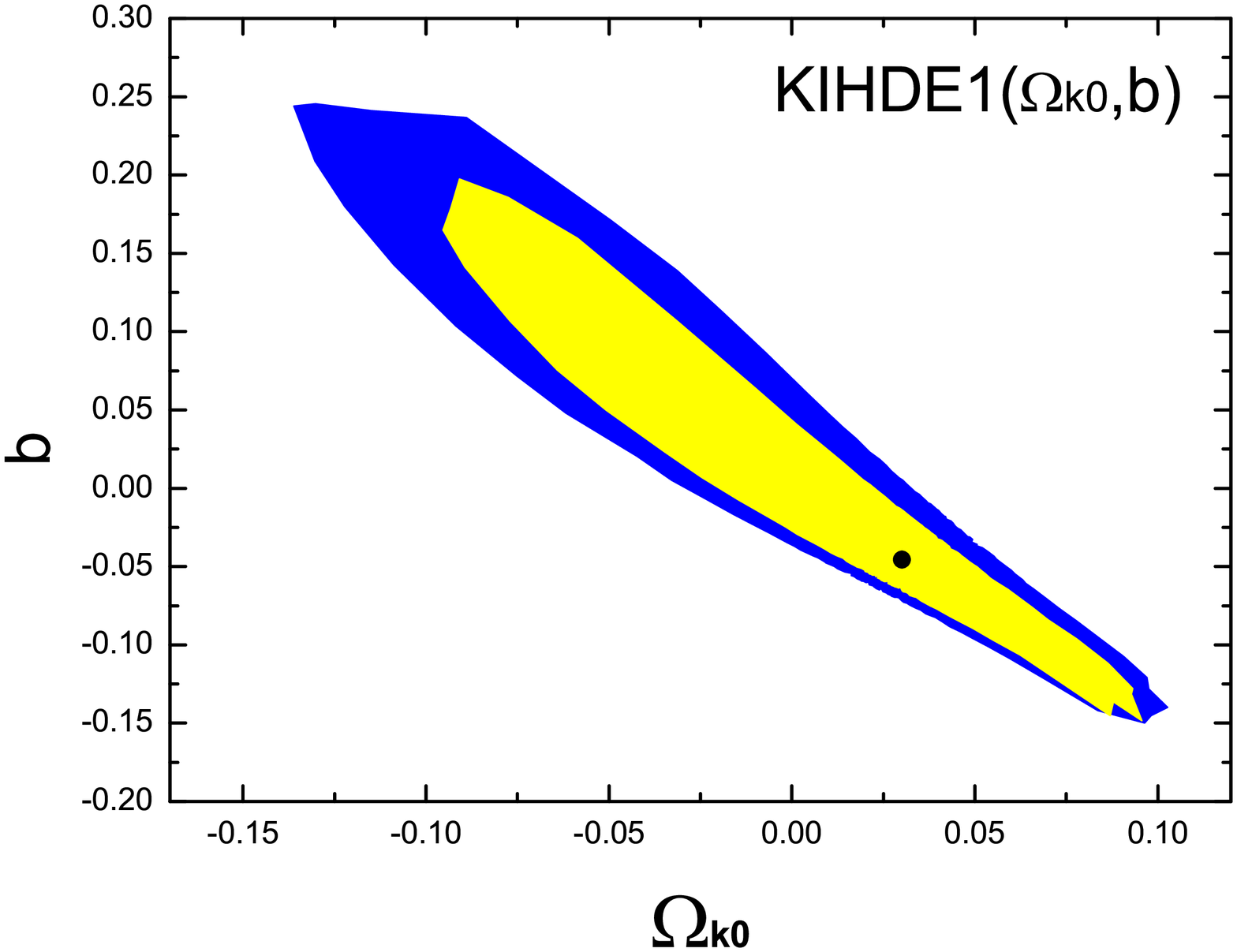}\hskip.5cm
\includegraphics[width=0.35\textwidth]{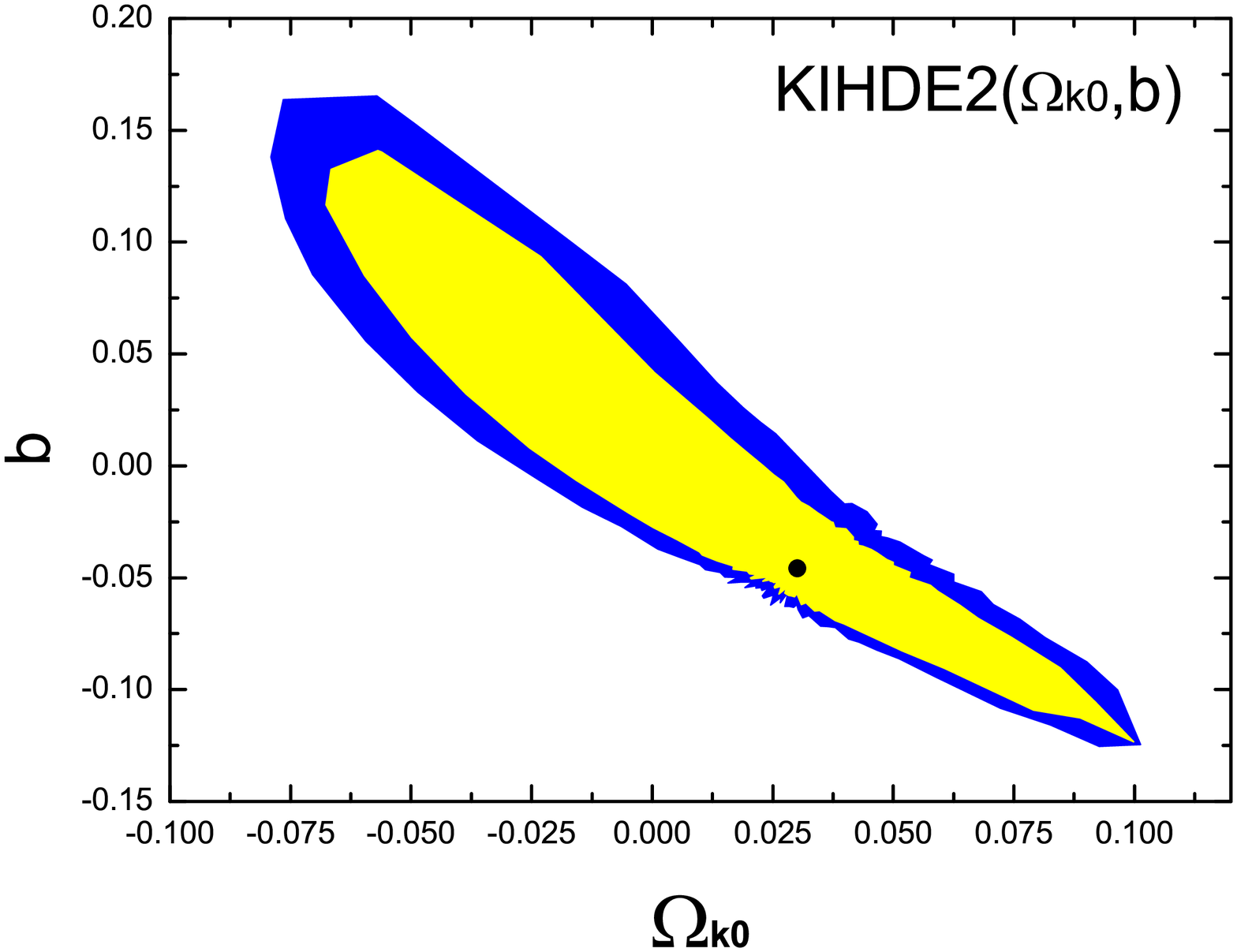}\vskip.5cm
\includegraphics[width=0.35\textwidth]{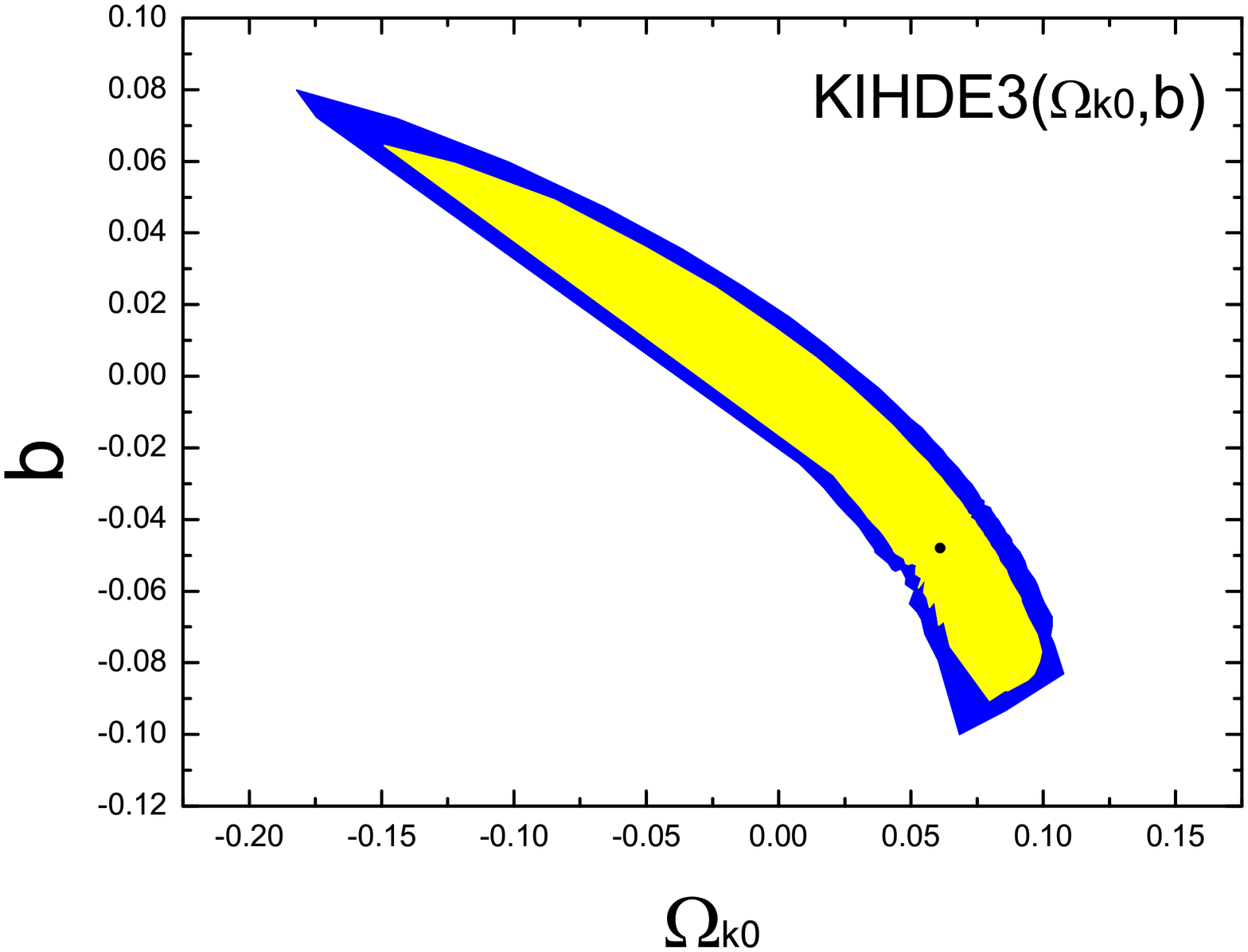}
\caption{Probability contours at $68.3\%$ and $95.4\%$ confidence
levels in the $\Omega_{k0}-b$ planes, for the three KIHDE
models.}\label{KIHDE}
\end{figure}

\subsection{Results and discussions}

Table \ref{table1} summarizes the fit results for the considered
models. In this table we show the best-fit and $1\sigma$ values of
the parameters and the $\chi^2_{min}$ of the models. Here,
$\Delta\chi^2_{min}=\chi^2_{min}({\rm HDE})-\chi^2_{min}$ stands for
the improvement of the model in $\chi^2_{min}$ comparing to the HDE
model. From this table, one can find two significant features. One
feature is that for the KHDE and IHDE models the values of
$\Delta\chi^2_{min}$ are rather small, and their best-fit parameters
(namely, $\Omega_{m0}$ and $c$) are very close to the corresponding
ones in the HDE model. The best-fit values and $1\sigma$ regions of
$\Omega_{k0}$ and $b$ are fairly small, implying that the
holographic dark energy model prefers zero curvature and zero
interaction. The other feature is that for the KIHDE models
$\Delta\chi^2_{min}$ as well as the best-fit and $1\sigma$ values of
$\Omega_{k0}$ and $b$ are much larger than those in the KHDE and
IHDE models. This indicates that there possibly exists some strong
degeneracy between the spatial curvature and the interaction in the
holographic dark energy model. Also, the best-fit values and
$1\sigma$ ranges of $c$ in the KIHDE models become remarkably larger
than those in other cases.

Figures \ref{fig-Omega_m0_and_c}$-$\ref{KIHDE} show the probability
contours at $68.3\%$ and $95.4\%$ confidence levels (CL) in various
parameter planes for these models. Let us discuss the various cases
in detail in what follows.

In Fig.~\ref{fig-Omega_m0_and_c}, we plot the $1\sigma$ and
$2\sigma$ contours in the $\Omega_{m0}-c$ plane for the KHDE and
IHDE models. We see from this figure that the inclusion of spatial
curvature or interaction in the holographic dark energy makes the
parameter space of $(\Omega_{m0}, c)$ become larger. For the HDE
model, we have $c<1$ in about $2\sigma$ range, for details see
Fig.~1 of Ref.~\cite{holoobs09}. However, when considering the
spatial curvature or the interaction, we can only obtain $c<1$ in
about $1\sigma$ range. In fact, in the IHDE models $c$ could be
evidently greater than one even in the $1\sigma$ range. This, in
some sense, is good news for the fate of the universe because the
likelihood of holographic dark energy becoming a phantom decreases.
The negative aspect lies in that the uncertainty is enlarged owing
to the amplification of the parameter space.

Figures~\ref{fig-Omega_m0_and_b} and \ref{fig-c_and_b} show the
degeneracy situations of $b$ and $\Omega_{m0}$, as well as $b$ and
$c$, in the IHDE models. It is clear that $b$ and $\Omega_{m0}$ are
in positive correlation, and $b$ and $c$ are anti-correlated. From
these two figures one can see that the observational data tell us
that $b$ can be both positive and negative. In our convention, a
positive $b$ will lead to an unphysical future for the universe
since $\rho_m$ will become negative and $\Omega_\Lambda$ will be
greater than one in the far future. So, one should only consider the
regions of $b\leq 0$ in the figures as realistic physical
situations. The best-fit values for $b$ are all close to zero,
although slightly larger than zero. Concretely, $b=6.1\times
10^{-5}$, $1.6\times 10^{-4}$ and $3.0\times 10^{-4}$, for the three
IHDE models, respectively. The $95\%$ CL limits on $b$ are fairly
small, in order of $10^{-2}$. This indicates that the observations
do not favor an interacting holographic dark energy model. However,
it should be pointed out that even a small $b$ could possibly be
used to avoid the future big rip caused by $c<1$. The sufficient and
necessary condition of avoiding the big rip in the interacting
holographic dark energy model with $c<1$ has been studied in detail
in Ref.~\cite{holoissue}. According to this work, we know that: for
IHDE1, the condition is $b\leq 1-c^{-2}$; for IHDE2, the condition
is $b\leq c^2-1$; for IHDE3, there is no late-time attractor
solution, so the big rip will be inevitable. From
Fig.~\ref{fig-c_and_b}, we find that for IHDE1 and IHDE2 the points
satisfying such sufficient and necessary conditions of avoiding the
big rip actually could be found in the $95\%$ CL region with $b<0$
and $c<1$.

The constraints on KHDE from the latest SNIa, CMB and BAO data are
shown in Fig.~\ref{KHDE}. The left panel shows that $\Omega_{k0}$
and $\Omega_{m0}$ are in positive correlation, and the right panel
indicates that $\Omega_{k0}$ and $c$ are anti-correlated. The
best-fit value for $\Omega_{k0}$ is $7.7\times 10^{-4}$, quite close
to zero. The values of $\Omega_{k0}$ in $2\sigma$ region can be both
positive and negative. The $95\%$ CL limit on $\Omega_{k0}$ is in
order of $10^{-2}$, nearly as good as that for a vacuum energy
model. These results indicate that the observations actually favor a
spatially flat holographic dark energy model.

Furthermore, we consider the most sophisticated case for the
holographic dark energy model, i.e., the interacting holographic
dark energy in a non-flat universe. Figure~\ref{KIHDE} shows the
constraints on $b$ and $\Omega_{k0}$ of the KIHDE models from the
latest SN, CMB and BAO data. We find that $b$ and $\Omega_{k0}$ are
strongly anti-correlated. It is remarkable that there exists
significant degeneracy between the phenomenological interaction and
spatial curvature in the holographic dark energy model. When
simultaneously considering the interaction and spatial curvature in
the holographic dark energy model, the parameter space is amplified,
especially, the ranges of $b$ and $\Omega_{k0}$ are enlarged by 10
times comparing to the IHDE and KHDE models. From Fig.~\ref{KIHDE},
we find that the holographic dark energy model with zero interaction
and flat geometry is still favored. Of course, the best-fit points
all indicate an interacting holographic dark energy (with $b$ about
$-0.05$) in a closed universe. When imposing the physical condition
$b<0$, one can find from Fig.~\ref{KIHDE} that a closed geometry is
more favored.

Finally, we feel that it would be better to make some additional
comments on the other related work of interacting holographic dark
energy models. In an early work \cite{Wang:2005jx}, the interacting
holographic dark energy model (namely, the case of IHDE2) was first
proposed; subsequently, this model was constrained by using the
observational data of that time \cite{Feng:2007wn}. Comparing with
these earlier studies, our results show a significant reduction of
the errors of parameters, owing to the more accurate data used. It
should also be mentioned that in Ref.~\cite{Wu:2008gg} the
interaction between holographic dark energy and matter was
reconstructed by employing two commonly used parametrization form
for the equation of state of dark energy. In addition, the
interacting holographic dark energy models with the Hubble scale as
IR cutoff were also discussed in, e.g.,
Refs.~\cite{Cruz:2008er,Xu:2009ys}. The interacting holographic dark
energy model in a spatially closed universe was investigated in
Ref.~\cite{MohseniSadjadi:2008na} where an additional requirement
that the entropy attributed to the IR cutoff $L$ should be
increasing was imposed. However, in the present paper, such a
condition is not imposed since an appropriate definition of entropy
for a holographic universe is still obscure for us (some studies
show that the Friedmann equation is consistent with the first law of
thermodynamics of the apparent horizon, see, e.g.,
Refs.~\cite{Cai:2005ra,Cai:2006rs}). It should also be noted that
there are some other versions of holographic dark energy model, for
example, the agegraphic dark energy model
\cite{Cai:2007us,newage,ageext} and the Ricci dark energy model
\cite{Gao:2007ep,ricciext}. The interacting agegraphic dark energy
model has been discussed \cite{Zhang:2009qa}, but the interacting
version of Ricci dark energy model has not been investigated.

\section{Conclusion}

In this paper we consider a sophisticated holographic dark energy
model where the interaction and spatial curvature are both involved.
The consideration of interaction between dark energy and matter in
the holographic dark energy is rather popular, since the interaction
not only can help to alleviate the cosmic coincidence problem but
also can be used to avoid the future big-rip singularity caused by
$c<1$. In addition, the consideration of spatial geometry in the
holographic dark energy is also necessary, because usually there
exists some degeneracy between the spatial curvature and the
dynamics of dark energy. The aim of this paper is to probe the
possible interaction and spatial curvature in the holographic dark
energy model in light of the latest observational data.

We considered three kinds of phenomenological interactions between
holographic dark energy and matter, i.e., the interaction term $Q$
is proportional to the energy densities of dark energy
($\rho_{\Lambda}$), matter ($\rho_m$) and dark energy plus matter
($\rho_\Lambda+\rho_m$). For the observational data, we used the
SNIa Constitution data, the shift parameter $R$ of the CMB from the
WMAP5, and the BAO parameter $A$ from the SDSS. We separately
considered the following cases: opening interaction but closing
spatial curvature; opening spatial curvature but closing
interaction; simultaneously opening interaction and spatial
curvature. Our results show that when separately considering the
interaction and spatial curvature, the interaction and spatial
curvature in the holographic dark energy model are both rather
small. In other words, the observations favor a non-interacting
holographic dark energy with flat geometry. When considering both
the interaction and the spatial curvature in the holographic dark
energy model, it is interesting to find that there exists remarkable
degeneracy between them. On the whole, according to the current
observational data, the holographic dark energy model without
interaction and spatial curvature is still more favored.

\begin{acknowledgments}
This work was supported by the Natural Science Foundation of China
under Grants Nos. 10705041, 10821504, 10975032 and 10975172, and
Ministry of Science and Technology 973 program under Grant No.
2007CB815401. SW also thanks the support from a graduate fund of the
University of Science and Technology of China.
\end{acknowledgments}



\begin{thebibliography}{}

\bibitem{Riess}
A.G. Riess  {\it et al.}, Astron.J. {\bf 116}, 1009  (1998); S.
Perlmutter {\it et al.}, ApJ {\bf 517}, 565 (1999); J. L. Tonry {\it
et al.}, ApJ {\bf 594}, 1 (2003); R.A. Knop {\it et al.}, ApJ {\bf
598}, 102 (2003); A.G. Riess {\it et al.}, ApJ {\bf 607}, 665
(2004).

\bibitem{spergel}  D.N. Spergel {\it et al.}, ApJS 148, 175 (2003);
C.L. Bennet {\it et al.}, ApJS. 148, 1 (2003); D.N. Spergel {\it et
al.}, ApJS {\bf 170}, 377 (2007); L. Page {\it et al.}, ApJS {\bf
170}, 335 (2007); G. Hinshaw {\it et al.}, ApJS {\bf 170}, 263
(2007).


\bibitem{Tegmark}
M. Tegmark {\it et al.}, Phys. Rev. D{\bf69}, 103501 (2004); ApJ
{\bf 606}, 702 (2004); Phys. Rev. D{\bf74}, 123507 (2006).

\bibitem{Weinberg} S. Weinberg, Rev. Mod. Phys. {\bf 61}, 1 (1989);
arXiv:astro-ph/0005265; V.~Sahni and A.A.~Starobinsky,
  Int.\ J.\ Mod.\ Phys.\  D {\bf 9}, 373 (2000); S.M. Carroll, Living Rev.Rel. {\bf 4}, 1
(2001); P.J.E. Peebles and B. Ratra, Rev. Mod. Phys. {\bf 75}, 559
(2003); T. Padmanabhan, Phys. Rept. {\bf 380}, 235 (2003); V.~Sahni,
  Lect.\ Notes Phys.\  {\bf 653}, 141 (2004);
E.~J.~Copeland, M.~Sami and S.~Tsujikawa,
  Int.\ J.\ Mod.\ Phys.\  D {\bf 15}, 1753 (2006);
  J.~Frieman, M.~Turner and D.~Huterer,
  Ann.\ Rev.\ Astron.\ Astrophys.\  {\bf 46}, 385 (2008).


\bibitem{quint} B. Ratra and P.J.E. Peebles,
                         Phys. Rev. D{\bf 37}, 3406 (1988);
                P.J.E. Peebles and B.Ratra,  ApJ {\bf325}, L17 (1988);
C. Wetterich, Nucl. Phys. B{\bf 302}, 668 (1988);
             A\&A {\bf 301}, 321 (1995);
R.R. Caldwell, R. Dave and P.J. Steinhardt,
       Phys. Rev. Lett. {\bf 80}, 1582 (1998);
I. Zlatev, L. Wang and P.J. Steinhardt
       Phys. Rev. Lett. {\bf 82}, 896 (1999).


\bibitem{phantom}
R.R. Caldwell, Phys. Lett. B {\bf 545}, 23 (2002); S.M. Carroll, M.
Hoffman and M. Trodden,
          Phys. Rev. D{\bf 68}, 023509 (2003);
R.R. Caldwell, M. Kamionkowski and N.N. Weinberg,
          Phys. Rev. Lett. {\bf 91},  071301 (2003).

\bibitem{k}
C. Armendariz-Picon, T. Damour and V. Mukhanov, Phys. Lett. B {\bf
458}, 209 (1999); C. Armendariz-Picon, V. Mukhanov and P.J.
Steinhardt, Phys. Rev. D{\bf 63}, 103510 (2001); T. Chiba, T. Okabe
and M. Yamaguchi, Phys. Rev. D{\bf 62}, 023511 (2000).


\bibitem{tachyonic} T. Padmanabhan, Phys. Rev. D{\bf 66}, 021301 (2002);
J.S. Bagla, H.K. Jassal, and T. Padmanabhan, Phys. Rev. D{\bf 67},
063504 (2003).

\bibitem{hessence} H. Wei, R.G. Cai, and D.F. Zeng, Class. Quant. Grav. {\bf 22}, 3189 (2005);
H. Wei, and R.G. Cai, Phys. Rev. D{\bf 72}, 123507 (2005); H. Wei,
N.N. Tang, and R.G. Cai, Phys. Rev. D{\bf 75}, 043009 (2007).

\bibitem{Chaplygin}
A.Y. Kamenshchik, U. Moschella and V. Pasquier, Phys. Lett. B {\bf
511}, 265 (2001); M.C. Bento, O. Bertolami and A.A. Sen, Phys. Rev.
D {\bf 66}, 043507 (2002); X. Zhang, F.Q. Wu and J. Zhang, JCAP {\bf
0601}, 003 (2006).


\bibitem{YMC} W.~Zhao and Y.~Zhang,
  Class.\ Quant.\ Grav.\  {\bf 23}, 3405 (2006);
T.Y. Xia and Y. Zhang, Phys. Lett. B {\bf 656}, 19 (2007); S. Wang,
Y. Zhang and T.Y. Xia, JCAP {\bf 10} 037 (2008); S. Wang and Y.
Zhang, Phys. Lett. B {\bf 669} 201(2008).

\bibitem{Witten:2000zk}
  E.~Witten,
  arXiv:hep-ph/0002297.



\bibitem{Holography} G. 't Hooft, gr-qc/9310026; L. Susskind, J. Math. Phys. \textbf{36}, 6377
(1995); J. D. Bekenstein, Phys. Rev. D \textbf{7}, 2333 (1973); J.
D. Bekenstein, Phys. Rev. D \textbf{9}, 3292 (1974); J. D.
Bekenstein, Phys. Rev. D \textbf{23}, 287 (1981); J. D. Bekenstein,
Phys. Rev. D \textbf{49}, 1912(1994); S. W. Hawking, Commun. Math.
Phys. \textbf{43}, 199 (1975); S. W. Hawking, Phys. Rev. D
\textbf{13}, 191 (1976).

\bibitem{cohen99} A. G. Cohen, D. B. Kaplan and A. E. Nelson, Phys. Rev. Lett. \textbf{82},
4971 (1999).

\bibitem{hsu04} S. D. H. Hsu, Phys. Lett. B \textbf{594}, 13 (2004).

\bibitem{li04} M. Li, Phys. Lett. B \textbf{603}, 1 (2004).

\bibitem{lmp} M. Li, M. X. Miao and Y. Pang, arXiv:0910.3375.

\bibitem{holode1}
  Q.~G.~Huang and M.~Li,
  JCAP {\bf 0408}, 013 (2004).

\bibitem{holode2}
  Q.~G.~Huang and M.~Li,
  JCAP {\bf 0503}, 001 (2005);
  X.~Zhang,
  Int.\ J.\ Mod.\ Phys.\  D {\bf 14}, 1597 (2005);
  Phys.\ Lett.\  B {\bf 648}, 1 (2007);
  Phys.\ Rev.\  D {\bf 74}, 103505 (2006);
B. Chen, M. Li and Y. Wang, Nucl. Phys. B {\bf 774}, 256 (2007);
J.F. Zhang, X. Zhang and H.Y. Liu, Eur. Phys. J. C {\bf 52}, 693
(2007); Phys.\ Lett.\  B {\bf 651}, 84 (2007); H. Wei and S.N.
Zhang, Phys. Rev. D {\bf 76}, 063003 (2007); C.J. Feng, Phys. Lett.
B {\bf 633}, 367 (2008); Y.~Z.~Ma and X.~Zhang,
  Phys.\ Lett.\  B {\bf 661}, 239 (2008); M. Li, X.D. Li, C.
Lin and Y. Wang, Commun. Theor. Phys. {\bf 51}, 181 (2009); Y.G.
Gong and T.J. Li, arXiv:0907.0860; 
  S.~Nojiri and S.~D.~Odintsov,
  Gen.\ Rel.\ Grav.\  {\bf 38}, 1285 (2006).


\bibitem{holodeobs}
  Q.~G.~Huang and Y.~G.~Gong,
  JCAP {\bf 0408}, 006 (2004);
  X.~Zhang and F.~Q.~Wu,
  Phys.\ Rev.\  D {\bf 72}, 043524 (2005);
  Phys.\ Rev.\  D {\bf 76}, 023502 (2007);
  Z.~Chang, F.~Q.~Wu and X.~Zhang,
  Phys.\ Lett.\  B {\bf 633}, 14 (2006);
  J.~Y.~Shen, B.~Wang, E.~Abdalla and R.~K.~Su,
  Phys.\ Lett.\  B {\bf 609}, 200 (2005);
  Z.~L.~Yi and T.~J.~Zhang,
  Mod.\ Phys.\ Lett.\  A {\bf 22}, 41 (2007);
  Y.~Z.~Ma, Y.~Gong and X. Chen,
  Eur.\ Phys.\ J.\  C {\bf 60}, 303 (2009).






\bibitem{holoobs09}
  M.~Li, X.~D.~Li, S.~Wang and X.~Zhang,
  JCAP {\bf 0906}, 036 (2009).

\bibitem{intholo}
  B.~Wang, C.~Y.~Lin and E.~Abdalla,
  Phys.\ Lett.\  B {\bf 637}, 357 (2006);
  B.~Wang, C.~Y.~Lin, D.~Pavon and E.~Abdalla,
  Phys.\ Lett.\  B {\bf 662}, 1 (2008);
  J.~Zhang, X.~Zhang and H.~Liu,
  Phys.\ Lett.\  B {\bf 659}, 26 (2008);
  H.~M.~Sadjadi and M.~Honardoost,
  Phys.\ Lett.\  B {\bf 647}, 231 (2007).

\bibitem{holoissue}
  M.~Li, C.~Lin and Y.~Wang,
  JCAP {\bf 0805}, 023 (2008).

\bibitem{healworld}
  X.~Zhang,
  arXiv:0909.4940.

\bibitem{Clarkson:2007bc}
  C.~Clarkson, M.~Cortes and B.~A.~Bassett,
  JCAP {\bf 0708}, 011 (2007).





\bibitem{Hicken2}
M.~Hicken {\it et al.},
  Astrophys.\ J.\  {\bf 700}, 1097 (2009).


\bibitem{Eisenstein}
D.J. Eisenstein {\it et al.}, ApJ. {\bf 633}, 560 (2005).

\bibitem{Komatsu}
E. Komatsu {\it et al.}, ApJS. {\bf 180}, 3301 (2009).

\bibitem{Bond}
J.R. Bond, G. Efstathiou and M. Tegmark, MNRAS. {\bf 291}, L33
(1997).

\bibitem{ywang3}
Y. Wang and P. Mukherjee, ApJ. {\bf 650}, 1 (2006).


\bibitem{Wang:2005jx}
  B.~Wang, Y.~Gong and E.~Abdalla,
  Phys.\ Lett.\  B {\bf 624}, 141 (2005).


\bibitem{Feng:2007wn}
  C.~Feng, B.~Wang, Y.~Gong and R.~K.~Su,
  JCAP {\bf 0709}, 005 (2007);
  Q.~Wu, Y.~Gong, A.~Wang and J.~S.~Alcaniz,
  Phys.\ Lett.\  B {\bf 659}, 34 (2008);
  K.~Karwan,
  JCAP {\bf 0805}, 011 (2008).


\bibitem{Wu:2008gg}
  S.~F.~Wu, P.~M.~Zhang and G.~H.~Yang,
  Class.\ Quant.\ Grav.\  {\bf 26}, 055020 (2009).

\bibitem{Cruz:2008er}
  N.~Cruz, S.~Lepe, F.~Pena and J.~Saavedra,
  Phys.\ Lett.\  B {\bf 669}, 271 (2008).

\bibitem{Xu:2009ys}
  L.~Xu,
  JCAP {\bf 0909}, 016 (2009).

\bibitem{MohseniSadjadi:2008na}
  H.~Mohseni Sadjadi and N.~Vadood,
  JCAP {\bf 0808}, 036 (2008).

\bibitem{Cai:2005ra}
  R.~G.~Cai and S.~P.~Kim,
  JHEP {\bf 0502}, 050 (2005).


\bibitem{Cai:2006rs}
  R.~G.~Cai and L.~M.~Cao,
  Phys.\ Rev.\  D {\bf 75}, 064008 (2007);
  R.~G.~Cai and L.~M.~Cao,
  Nucl.\ Phys.\  B {\bf 785}, 135 (2007);
  Y.~Gong and A.~Wang,
  Phys.\ Rev.\ Lett.\  {\bf 99}, 211301 (2007);
  Y.~Gong, B.~Wang and A.~Wang,
  JCAP {\bf 0701}, 024 (2007);
  R.~G.~Cai, L.~M.~Cao and Y.~P.~Hu,
  Class.\ Quant.\ Grav.\  {\bf 26}, 155018 (2009).

\bibitem{Cai:2007us}
  R.~G.~Cai,
  Phys.\ Lett.\  B {\bf 657}, 228 (2007).

\bibitem{newage}
  H.~Wei and R.~G.~Cai,
  Phys.\ Lett.\  B {\bf 660}, 113 (2008).

\bibitem{ageext}
  X.~Wu, Y.~Zhang, H.~Li, R.~G.~Cai and Z.~H.~Zhu,
  arXiv:0708.0349;
  H.~Wei and R.~G.~Cai,
  Phys.\ Lett.\  B {\bf 663}, 1 (2008);
  I.~P.~Neupane,
  Phys.\ Rev.\  D {\bf 76}, 123006 (2007);
  J.~Zhang, X.~Zhang and H.~Liu,
  Eur.\ Phys.\ J.\  C {\bf 54}, 303 (2008);
  Y.~W.~Kim, H.~W.~Lee, Y.~S.~Myung and M.~I.~Park,
  Mod.\ Phys.\ Lett.\  A {\bf 23}, 3049 (2008).
  J.~P.~Wu, D.~Z.~Ma and Y.~Ling,
  Phys.\ Lett.\  B {\bf 663}, 152 (2008);
  J.~Cui, L.~Zhang, J.~Zhang and X.~Zhang,
  arXiv:0902.0716;
  X.~L.~Liu and X.~Zhang,
  Commun.\ Theor.\ Phys.\  {\bf 52}, 761 (2009).

\bibitem{Gao:2007ep}
  C.~Gao, F. Q. Wu, X.~Chen and Y.~G.~Shen,
  Phys. Rev. D {\bf 79}, 043511 (2009).

\bibitem{ricciext}
  C.~J.~Feng,
  arXiv:0806.0673;
  C.~J.~Feng,
  Phys.\ Lett.\  B {\bf 670}, 231 (2008);
  C.~J.~Feng,
  Phys.\ Lett.\  B {\bf 672}, 94 (2009);
  X.~Zhang,
  Phys.\ Rev.\  D {\bf 79}, 103509 (2009);
  L.~Xu, W.~Li and J.~Lu,
  arXiv:0810.4730;
  C.~J.~Feng,
  arXiv:0812.2067;
  K.~Y.~Kim, H.~W.~Lee and Y.~S.~Myung,
  arXiv:0812.4098.
  R.~G.~Cai, B.~Hu and Y.~Zhang,
  Commun. Theor. Phys. {\bf 51}, 954 (2009).
  C.~J.~Feng and X.~Zhang,
  Phys.\ Lett.\  B {\bf 680}, 399 (2009);
  L.~N.~Granda, W.~Cardona and A.~Oliveros,
  arXiv:0910.0778.


\bibitem{Zhang:2009qa}
  L.~Zhang, J.~Cui, J.~Zhang and X.~Zhang,
  arXiv:0911.2838;
  A.~Sheykhi,
  Phys.\ Lett.\  B {\bf 680}, 113 (2009).











\end{thebibliography}
\end{document}